\documentclass[letterpaper,DIV=12,abstracton]{scrartcl}

\usepackage[utf8]{inputenc}
\usepackage{hyperref}
\usepackage{graphicx}
\usepackage{xcolor}

\usepackage{amsmath}
\usepackage{amsthm}
\usepackage{amssymb}
\usepackage{amsfonts}
\usepackage{bbm}

\usepackage{IEEEtrantools}
\allowdisplaybreaks

\usepackage[maxbibnames=9,maxcitenames=9]{biblatex}
\usepackage{enumitem}

\usepackage{tikz}
\usetikzlibrary{calc}
\usetikzlibrary{arrows}
\usetikzlibrary{arrows.meta}
\usetikzlibrary{patterns}
\usetikzlibrary{positioning}
\usetikzlibrary{decorations.pathreplacing}
\usetikzlibrary{shapes.misc}
\usetikzlibrary{spy}

\usepackage{pgfplots}
\pgfplotsset{compat=1.14}

\definecolor{myParula01Blue}{RGB}{0,114,189}
\definecolor{myParula02Orange}{RGB}{217,83,25}
\definecolor{myParula03Yellow}{RGB}{237,177,32}
\definecolor{myParula04Purple}{RGB}{126,47,142}
\definecolor{myParula05Green}{RGB}{119,172,48}
\definecolor{myParula06LightBlue}{RGB}{77,190,238}
\definecolor{myParula07Red}{RGB}{162,20,47}

\tikzset{myparula11/.style={color=myParula01Blue,solid,mark=+,mark options={solid}}}
\tikzset{myparula12/.style={color=myParula01Blue,densely dashed,mark=x,mark options={solid}}}
\tikzset{myparula13/.style={color=myParula01Blue,densely dotted,mark=o,mark options={solid}}}
\tikzset{myparula14/.style={color=myParula01Blue,dashdotted,mark=triangle,mark options={solid}}}
\tikzset{myparula15/.style={color=myParula01Blue,dashdotdotted,mark=square,mark options={solid}}}

\tikzset{myparula21/.style={color=myParula02Orange,solid,mark=+,mark options={solid}}}
\tikzset{myparula22/.style={color=myParula02Orange,densely dashed,mark=x,mark options={solid}}}
\tikzset{myparula23/.style={color=myParula02Orange,densely dotted,mark=o,mark options={solid}}}
\tikzset{myparula24/.style={color=myParula02Orange,dashdotted,mark=triangle,mark options={solid}}}
\tikzset{myparula25/.style={color=myParula02Orange,dashdotdotted,mark=square,mark options={solid}}}

\tikzset{myparula31/.style={color=myParula03Yellow,solid,mark=+,mark options={solid}}}
\tikzset{myparula32/.style={color=myParula03Yellow,densely dashed,mark=x,mark options={solid}}}
\tikzset{myparula33/.style={color=myParula03Yellow,densely dotted,mark=o,mark options={solid}}}
\tikzset{myparula34/.style={color=myParula03Yellow,dashdotted,mark=triangle,mark options={solid}}}
\tikzset{myparula35/.style={color=myParula03Yellow,dashdotdotted,mark=square,mark options={solid}}}

\tikzset{myparula41/.style={color=myParula04Purple,solid,mark=+,mark options={solid}}}
\tikzset{myparula42/.style={color=myParula04Purple,densely dashed,mark=x,mark options={solid}}}
\tikzset{myparula43/.style={color=myParula04Purple,densely dotted,mark=o,mark options={solid}}}
\tikzset{myparula44/.style={color=myParula04Purple,dashdotted,mark=triangle,mark options={solid}}}
\tikzset{myparula45/.style={color=myParula04Purple,dashdotdotted,mark=square,mark options={solid}}}

\tikzset{myparula51/.style={color=myParula05Green,solid,mark=+,mark options={solid}}}
\tikzset{myparula52/.style={color=myParula05Green,densely dashed,mark=x,mark options={solid}}}
\tikzset{myparula53/.style={color=myParula05Green,densely dotted,mark=o,mark options={solid}}}
\tikzset{myparula54/.style={color=myParula05Green,dashdotted,mark=triangle,mark options={solid}}}
\tikzset{myparula55/.style={color=myParula05Green,dashdotdotted,mark=square,mark options={solid}}}

\tikzset{myparula61/.style={color=myParula06LightBlue,solid,mark=+,mark options={solid}}}
\tikzset{myparula62/.style={color=myParula06LightBlue,densely dashed,mark=x,mark options={solid}}}
\tikzset{myparula63/.style={color=myParula06LightBlue,densely dotted,mark=o,mark options={solid}}}
\tikzset{myparula64/.style={color=myParula06LightBlue,dashdotted,mark=triangle,mark options={solid}}}
\tikzset{myparula65/.style={color=myParula06LightBlue,dashdotdotted,mark=square,mark options={solid}}}

\tikzset{myparula71/.style={color=myParula07Red,solid,mark=+,mark options={solid}}}
\tikzset{myparula72/.style={color=myParula07Red,densely dashed,mark=x,mark options={solid}}}
\tikzset{myparula73/.style={color=myParula07Red,densely dotted,mark=o,mark options={solid}}}
\tikzset{myparula74/.style={color=myParula07Red,dashdotted,mark=triangle,mark options={solid}}}
\tikzset{myparula75/.style={color=myParula07Red,dashdotdotted,mark=square,mark options={solid}}}

\usepackage{xspace}
\newcommand{\cf}[0]{cf.\xspace}
\newcommand{\ie}[0]{\emph{i.e.}\xspace}

\tikzset{blockchain/.style={
        x=1.25cm,
        y=1.25cm,
        node distance=0.5cm,
        block/.style = {
            minimum width=0.75cm,
            minimum height=0.75cm,
            draw,
            shade,
            top color=white,
            bottom color=black!10,
        },
        block-adv/.style = {
            block,
            bottom color=myParula07Red!50,
            draw=myParula07Red!50!black,
        },
        block-hon/.style = {
            block,
            bottom color=myParula05Green!50,
            draw=myParula05Green!50!black,
        },
        link/.style = {
            -latex,
        },
        link-adv/.style = {
            link,
        },
        link-hon/.style = {
            link,
        },
    }
}

\newcommand{\Wp}[0]{\ensuremath{W_{\mathrm{p}}}}
\newcommand{\Tconf}[0]{\ensuremath{T_\mathsf{conf}}}
\newcommand{\tx}[0]{\ensuremath{\mathsf{tx}}}
\newcommand{\LOG}[2]{%
    \ensuremath{\mathsf{Ledger}_{#1}^{#2}}
}

\title{Two Attacks On Proof-of-Stake GHOST/Ethereum}
\author{%
    Joachim Neu\\%
    {\small\texttt{jneu@stanford.edu}}%
    \and%
    Ertem Nusret Tas\\%
    {\small\texttt{nusret@stanford.edu}}%
    \and%
    David Tse\\%
    {\small\texttt{dntse@stanford.edu}}}
\date{March 2, 2022}

\begin{document}

\maketitle
\begin{abstract}
We present two attacks targeting the
Proof-of-Stake (PoS) Ethereum
consensus protocol.
The first attack 
suggests a fundamental conceptual incompatibility between
PoS
and the
Greedy Heaviest-Observed Sub-Tree (GHOST)
fork choice paradigm employed by PoS Ethereum.
In a nutshell,
PoS allows an adversary with
a vanishing amount of stake to produce an unlimited
number of equivocating blocks.
While most equivocating blocks will be orphaned,
such orphaned `uncle blocks' still influence fork choice
under the GHOST paradigm, bestowing upon the adversary
devastating control over the canonical chain.
While the Latest Message Driven (LMD) aspect of current PoS Ethereum
prevents a straightforward application of this attack,
our second attack
shows how LMD specifically can be exploited
to obtain a new variant of the balancing
attack that overcomes
a recent protocol addition that was intended to
mitigate balancing-type attacks.
Thus, in its current form,
PoS Ethereum without and with LMD
is vulnerable to our first and second attack, respectively.
\end{abstract}
{\let\thefootnote\relax\footnote{JN and ENT contributed equally and are listed alphabetically.}}

\section{Introduction}
\label{sec:introduction}

The currently proposed Proof-of-Stake (PoS) Ethereum consensus protocol \cite{eth2-spec-beaconchain,eth2-spec-forkchoice,eth2-spec-validator}
is constructed from an application of the finality gadget
Casper FFG \cite{casper} on top of the fork choice rule LMD GHOST,
a variant of the Greedy Heaviest-Observed Sub-Tree (GHOST) \cite{ghost} rule
which considers only each participant's most recent vote
(Latest Message Driven, LMD).
Subsequently, we refer
as \emph{validators}
to participants with stake that allows them to vote as part of the protocol.
A slightly simplified and analytically more tractable variant of
the proposed PoS Ethereum protocol
is given by the Gasper protocol \cite{gasper}.
The protocol is recapitulated on a high level in Section~\ref{sec:modelprotocol}.

This report continues a sequence of earlier works
\cite{ebbandflow,ethresearch-balancing-attack,ethresearch-balancing-attack2,3attacks,lowcostreorgs,ethresearch-bouncing-attack,ethresearch-bouncing-attack-analysis,ethresearch-bouncing-attack-prevention,aadilemma}
that highlighted 
security vulnerabilities in PoS Ethereum and the constituent protocols Casper FFG and LMD GHOST.
More specifically,
we report two new attacks.

The first attack, called \emph{avalanche attack} and described
in Section~\ref{sec:avalanche-attack} (\cf \cite{ethresearch-avalanche-attack}),
suggests a fundamental conceptual incompatibility between
PoS on the one hand
and the GHOST fork choice paradigm on the other hand.
In a nutshell,
`orphaned` so called `uncle blocks' off the canonical chain
still influence the fork choice in GHOST.
In Proof-of-Work (PoW), for which GHOST is secure
\cite{aggelosghost}, this does not create a problem
because the number of uncle blocks which the adversary
controls and by which it can influence the fork choice
is bounded by the PoW mechanism.
PoS, on the other hand, allows the adversary to equivocate,
\ie, for each block production opportunity the adversary
can produce an unlimited amount of equivocating blocks
extending different parent blocks of the block tree.
As a result,
GHOST's way of accounting for uncle blocks
in fork choice gives the adversary much more influence
over fork choice in PoS than in PoW,
casting doubt over whether GHOST should be used with PoS at all.

The LMD aspect of current PoS Ethereum
interferes with a naive application of the avalanche attack
to PoS Ethereum.
However, our second attack,
described in Section~\ref{sec:attack-lmd-ghost} (\cf \cite{ethresearch-balancing-attack-lmd}),
shows how the LMD feature specifically can be exploited
to obtain a new variant of the balancing
attack \cite{ethresearch-balancing-attack,ethresearch-balancing-attack2}.
This new attack also overcomes
`proposer boosting' \cite{mitigationlmdghostbalancingattacks},
a recent protocol addition that was intended to
mitigate balancing-type attacks.

Thus, in its current form,
PoS Ethereum without LMD is vulnerable to our first attack
(avalanche attack), and PoS Ethereum with LMD is vulnerable
to our second attack
(new LMD-specific variant of earlier balancing attack).

\section{Proof-of-Stake Ethereum Consensus Protocol}
\label{sec:modelprotocol}

Let $\LOG{i}{t}$ denote the transaction ledger output by honest
validator $i$ at time $t$.
\emph{Security} of a consensus protocol such as PoS Ethereum is comprised of
\emph{safety} and
\emph{liveness}:
\begin{itemize}
    \item \textbf{Safety:} For any given times $t,t'$ and honest validators $i,j$, either $\LOG{i}{t}$ is a prefix of $\LOG{j}{t'}$, or vice versa. %
    Less formally, the ledgers output by two honest validators at two points in time are consistent with each other.

    \item \textbf{Liveness:} If a transaction $\tx$ is received by all honest validators by some time $t$, then $\tx$ appears 
    in $\LOG{i}{t'}$ for any time $t' \geq t+\Tconf$ and for any honest validator $i$, where $\Tconf$ is the protocol's \emph{confirmation time}.
    Less formally,
    transactions get confirmed in honest validators' ledgers with at most $\Tconf$ time delay.
\end{itemize}

\label{sec:modelprotocol-protocol}

We first describe a simplified 
version of PoS Ethereum without the LMD rule,
reduced to the relevant fork choice mechanics.
Although we present a self-contained description, familiarity with the protocols GHOST \cite{ghost} and Gasper \cite{gasper}, as well as with the beacon chain's fork choice specification \cite{eth2-spec-beaconchain} and proposal weights\footnote{\url{https://github.com/ethereum/consensus-specs/pull/2730}} \cite{mitigationlmdghostbalancingattacks} are useful for a deeper understanding of this section.

In the consensus protocol (also called Committee-GHOST),
time proceeds in synchronized slots of duration $2\Delta$,
since
it is assumed that message delay between honest validators
is bounded by $\Delta$.
For each slot, one \emph{proposer}
and a \emph{committee} of $W$ validators
are drawn independently and uniformly at random 
(without replacement for the committee members)
from the $N$ validators.
We say a slot is honest or adversarial if the corresponding block proposer is honest or adversarial, respectively.
The following fork-choice rule is used
in the view of validator $i$ at slot $t$
to determine a canonical block
and its prefix of blocks as the \emph{canonical chain}:
\begin{itemize}
\item
Starting at the genesis block $b_0$, sum for each child block $b$
the number of unique valid votes for that block and its descendents.
Here, unique means counting
only one vote per committee member in each slot $s<t$.
Valid means that the vote from a committee member of slot $s<t$ was cast for a block produced in a slot $s'\leq s$.
Ties are broken by the adversary.

Add a proposal boost of $\Wp$ to the score of $b$
if $b$ or one of its descendants is a valid
proposal from the current slot $t$.
Here, valid means that the block is not from a future slot,
was produced by the proposer of slot $t$,
and that proposal time slots along block chains are strictly increasing.

\item
Pick the child block $b^*$ with highest score (\cf GHOST rule \cite{ghost}),
breaking ties adversarially.

\item
Recurse ($b_0 \leftarrow b^*$) until reaching a leaf block.
Output that block.
\end{itemize}

At the beginning of each slot, the slot's proposer determines a block using the fork-choice rule and extends it with a new proposal.
Half way into each slot (\ie, $\Delta$ time after the proposal and after the
beginning of the slot), the slot's committee members determine a block using the fork-choice rule in their local view and vote for it.
The unit of time is a time slot.
A block from slot $t$ and its prefix are confirmed
and output as the ledger
if and only if at time $t+\Tconf + \frac{1}{2}$ (\ie, at the time of voting in slot $t+\Tconf$),
the block is in the chain determined by the fork-choice rule
in the view of the respective honest validator.
Here, $\Tconf$ is the \emph{confirmation time}.

For simplicity, we assume
a fraction $\beta$ such that
(a) for any given slot
the probability of the block proposer being adversarial
is at most $\beta$,
and (b) the fraction of adversarial validators in any committee
is at most $\beta$ throughout the protocol's execution.
\label{sec:modelprotocol-protocol-lmd}

The LMD rule modifies the above protocol
as follows.
Every validator keeps a table of `latest votes' received from the other validators, in the following manner: 
When a valid vote from a validator is received, then the table entry for that validator is updated, if and only if the new vote is from a slot \emph{strictly later than} the current entry. 
Hence, when a validator observes a slot-$t$ vote from the committee member $i$ of some slot $t$, it records this vote and ignores all subsequent slot-$t'$ votes for $t' \leq t$ by $i$.
Thus,
if a validator receives two equivocating votes from the same validator for the \emph{same} time slot, the validator counts only the vote received earlier in time.

\section{Avalanche Attack on Proof-of-Stake GHOST}
\label{sec:avalanche-attack}

We describe a generic attack on PoS GHOST variants. 
This points to conceptual issues with the combination of PoS and GHOST,
awareness of which might be of interest beyond PoS Ethereum fork choice design. 
PoS Ethereum, as it stands, is not susceptible to this attack (due to LMD, which comes with its own problems, see Section~\ref{sec:attack-lmd-ghost}). 

We assume basic familiarity with GHOST \cite{ghost} and Gasper \cite{gasper}.
For details, we refer the interested reader
to the beacon chain's fork choice specification \cite{eth2-spec-beaconchain} and earlier attacks \cite{ethresearch-balancing-attack,ethresearch-balancing-attack2}.

\subsection{High Level Description}
\label{sec:avalanche-high-level}

The \emph{avalanche attack} on PoS GHOST combines \emph{selfish mining} \cite{selfishmining} with \emph{equivocations}.
The adversary uses withheld blocks to displace an honest chain once it catches up in sub-tree weight with the number of withheld adversarial blocks. 
The withheld blocks are released in a flat but wide sub-tree, exploiting the fact that under the GHOST rule such a sub-tree can displace a long chain. 
Only two withheld blocks enter the canonical chain permanently, while the other withheld blocks are subsequently reused through equivocations to build further sub-trees to displace even more honest blocks. 
The attack exploits a specific weakness of the GHOST rule in combination with equivocations, namely that an adversary can reuse `uncle blocks' in GHOST, and thus such equivocations contribute to the weight of multiple ancestors. 
This casts doubt over whether GHOST should be used with PoS at all.

We also provide a proof-of-concept implementation for vanilla PoS GHOST and Committee-GHOST.\footnote{Source code: \url{https://github.com/tse-group/pos-ghost-attack}} 
By `vanilla PoS GHOST', we mean a one-to-one translation of GHOST
\cite{ghost} from proof-of-work lotteries to proof-of-stake lotteries. 
In that case, every block comes with unit weight.
By `Committee-GHOST' we mean a vote-based variant of GHOST as used in PoS Ethereum, where block weight is determined by votes and potentially a proposal boost \cite{mitigationlmdghostbalancingattacks}. 
Subsequently, we first illustrate the attack with an example, then provide a more detailed description, and finally show plots produced by the proof-of-concept implementation.

\subsection{A Simple Attack Example}
\label{sec:avalanche-simple-attack}

\begin{figure}
    \centering
    \begin{tikzpicture}[blockchain,scale=0.9]
    
        \node [block] (G) at (0,0) {G};
        
        \node [block-hon] (h1) at (0,-1) {1};
        \node [block-hon] (h2) at (0,-2) {2};
        \node [block-hon] (h3) at (0,-3) {3};
        \node [block-hon] (h4) at (0,-4) {4};
        
        \draw [link-hon] (h4) -- (h3);
        \draw [link-hon] (h3) -- (h2);
        \draw [link-hon] (h2) -- (h1);
        \draw [link-hon] (h1) -- (G);
        
        \node [block-adv,opacity=0.3] (a1) at (1,-1) {1};
        \node [block-adv,opacity=0.3] (a2) at (1,-2) {2};
        \node [block-adv,opacity=0.3] (a3) at (2,-2) {3};
        \node [block-adv,opacity=0.3] (a4) at (3,-2) {4};
        \node [block-adv,opacity=0.3] (a5) at (4,-2) {5};
        \node [block-adv,opacity=0.3] (a6) at (5,-2) {6};
        
        \draw [link-adv,opacity=0.3] (a3.north) -- (a1);
        \draw [link-adv,opacity=0.3] (a4.north) -- (a1);
        \draw [link-adv,opacity=0.3] (a5.north) -- (a1);
        \draw [link-adv,opacity=0.3] (a6.north) -- (a1);
        \draw [link-adv,opacity=0.3] (a2.north) -- (a1);
        \draw [link-adv,opacity=0.3] (a1.north) -- (G);
        
        \node [block-hon,opacity=0.3] (hNEW) at (0,-5) {...};
        \draw [link-hon,opacity=0.3] (hNEW) -- (h4);
    
    \end{tikzpicture}
    \caption{First, the adversary withholds its flat-but-wide sub-tree of $k=6$ withheld blocks, while honest nodes produce a chain. (Green and red indicate honest and adversarial blocks respectively, and the numbers on blocks indicate which block production opportunity of honest/adversary they correspond to.)}
    \label{fig:avalanche-attack-1}
\end{figure}
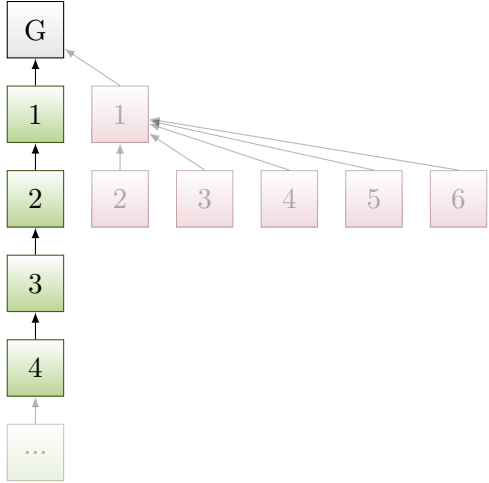

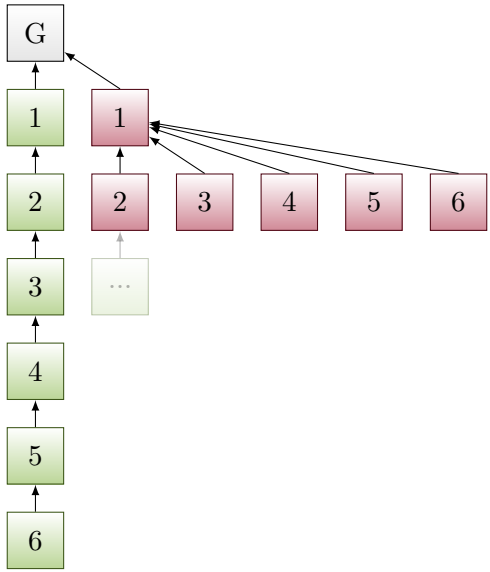
\begin{figure}
    \centering
    \begin{tikzpicture}[blockchain,scale=0.9]
    
        \node [block] (G) at (0,0) {G};
        
        \node [block-hon] (h1) at (0,-1) {1};
        \node [block-hon] (h2) at (0,-2) {2};
        \node [block-hon] (h3) at (0,-3) {3};
        \node [block-hon] (h4) at (0,-4) {4};
        \node [block-hon] (h5) at (0,-5) {5};
        \node [block-hon] (h6) at (0,-6) {6};
        
        \draw [link-hon] (h6) -- (h5);
        \draw [link-hon] (h5) -- (h4);
        \draw [link-hon] (h4) -- (h3);
        \draw [link-hon] (h3) -- (h2);
        \draw [link-hon] (h2) -- (h1);
        \draw [link-hon] (h1) -- (G);
        
        \node [block-adv] (a1) at (1,-1) {1};
        \node [block-adv] (a2) at (1,-2) {2};
        \node [block-adv] (a3) at (2,-2) {3};
        \node [block-adv] (a4) at (3,-2) {4};
        \node [block-adv] (a5) at (4,-2) {5};
        \node [block-adv] (a6) at (5,-2) {6};
        
        \draw [link-adv] (a3.north) -- (a1);
        \draw [link-adv] (a4.north) -- (a1);
        \draw [link-adv] (a5.north) -- (a1);
        \draw [link-adv] (a6.north) -- (a1);
        \draw [link-adv] (a2.north) -- (a1);
        \draw [link-adv] (a1.north) -- (G);
        
        \node [block-hon,opacity=0.3] (hNEW) at (1,-3) {...};
        \draw [link-hon,opacity=0.3] (hNEW) -- (a2);
    
    \end{tikzpicture}
    \caption{Once honest nodes build a chain of length $k=6$, the adversary releases the withheld blocks, and displaces the honest chain.}
    \label{fig:avalanche-attack-2}
\end{figure}

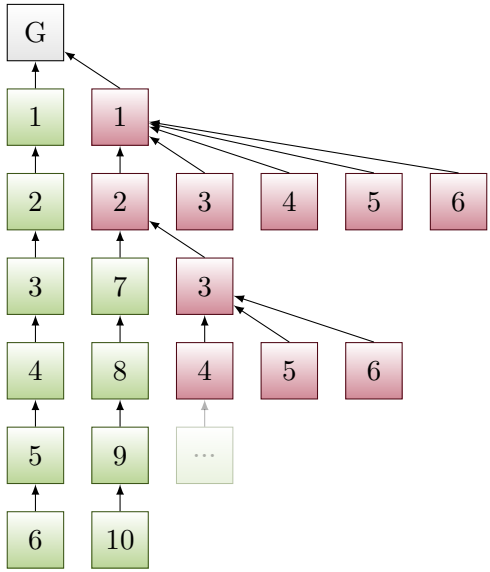
\begin{figure}
    \centering
    \begin{tikzpicture}[blockchain,scale=0.9]
    
        \node [block] (G) at (0,0) {G};
        
        \node [block-hon] (h1) at (0,-1) {1};
        \node [block-hon] (h2) at (0,-2) {2};
        \node [block-hon] (h3) at (0,-3) {3};
        \node [block-hon] (h4) at (0,-4) {4};
        \node [block-hon] (h5) at (0,-5) {5};
        \node [block-hon] (h6) at (0,-6) {6};
        
        \draw [link-hon] (h6) -- (h5);
        \draw [link-hon] (h5) -- (h4);
        \draw [link-hon] (h4) -- (h3);
        \draw [link-hon] (h3) -- (h2);
        \draw [link-hon] (h2) -- (h1);
        \draw [link-hon] (h1) -- (G);
        
        \node [block-adv] (a1) at (1,-1) {1};
        \node [block-adv] (a2) at (1,-2) {2};
        \node [block-adv] (a3) at (2,-2) {3};
        \node [block-adv] (a4) at (3,-2) {4};
        \node [block-adv] (a5) at (4,-2) {5};
        \node [block-adv] (a6) at (5,-2) {6};
        
        \draw [link-adv] (a3.north) -- (a1);
        \draw [link-adv] (a4.north) -- (a1);
        \draw [link-adv] (a5.north) -- (a1);
        \draw [link-adv] (a6.north) -- (a1);
        \draw [link-adv] (a2.north) -- (a1);
        \draw [link-adv] (a1.north) -- (G);
        
        \node [block-hon] (h7) at (1,-3) {7};
        \node [block-hon] (h8) at (1,-4) {8};
        \node [block-hon] (h9) at (1,-5) {9};
        \node [block-hon] (h10) at (1,-6) {10};
        
        \draw [link-hon] (h10) -- (h9);
        \draw [link-hon] (h9) -- (h8);
        \draw [link-hon] (h8) -- (h7);
        \draw [link-hon] (h7) -- (a2);
        
        \node [block-adv] (a3) at (2,-3) {3};
        \node [block-adv] (a4) at (2,-4) {4};
        \node [block-adv] (a5) at (3,-4) {5};
        \node [block-adv] (a6) at (4,-4) {6};
        
        \draw [link-adv] (a5.north) -- (a3);
        \draw [link-adv] (a6.north) -- (a3);
        \draw [link-adv] (a4.north) -- (a3);
        \draw [link-adv] (a3.north) -- (a2);
        
        \node [block-hon,opacity=0.3] (hNEW) at (2,-5) {...};
        \draw [link-hon,opacity=0.3] (hNEW) -- (a4);
    
    \end{tikzpicture}
    \caption{Note that the adversary can reuse its blocks $3, 4, 5, 6$. Honest nodes build a new chain on top of $2 \rightarrow 1 \rightarrow \mathsf{Genesis}$. Once that new chain reaches length $4$, the adversary releases another displacing sub-tree.}
    \label{fig:avalanche-attack-3}
\end{figure}

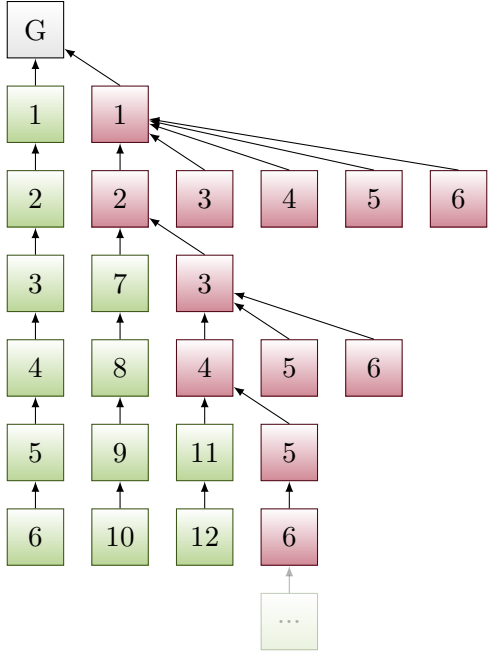
\begin{figure}
    \centering
    \begin{tikzpicture}[blockchain,scale=0.9]
    
        \node [block] (G) at (0,0) {G};
        
        \node [block-hon] (h1) at (0,-1) {1};
        \node [block-hon] (h2) at (0,-2) {2};
        \node [block-hon] (h3) at (0,-3) {3};
        \node [block-hon] (h4) at (0,-4) {4};
        \node [block-hon] (h5) at (0,-5) {5};
        \node [block-hon] (h6) at (0,-6) {6};
        
        \draw [link-hon] (h6) -- (h5);
        \draw [link-hon] (h5) -- (h4);
        \draw [link-hon] (h4) -- (h3);
        \draw [link-hon] (h3) -- (h2);
        \draw [link-hon] (h2) -- (h1);
        \draw [link-hon] (h1) -- (G);
        
        \node [block-adv] (a1) at (1,-1) {1};
        \node [block-adv] (a2) at (1,-2) {2};
        \node [block-adv] (a3) at (2,-2) {3};
        \node [block-adv] (a4) at (3,-2) {4};
        \node [block-adv] (a5) at (4,-2) {5};
        \node [block-adv] (a6) at (5,-2) {6};
        
        \draw [link-adv] (a3.north) -- (a1);
        \draw [link-adv] (a4.north) -- (a1);
        \draw [link-adv] (a5.north) -- (a1);
        \draw [link-adv] (a6.north) -- (a1);
        \draw [link-adv] (a2.north) -- (a1);
        \draw [link-adv] (a1.north) -- (G);
        
        \node [block-hon] (h7) at (1,-3) {7};
        \node [block-hon] (h8) at (1,-4) {8};
        \node [block-hon] (h9) at (1,-5) {9};
        \node [block-hon] (h10) at (1,-6) {10};
        
        \draw [link-hon] (h10) -- (h9);
        \draw [link-hon] (h9) -- (h8);
        \draw [link-hon] (h8) -- (h7);
        \draw [link-hon] (h7) -- (a2);
        
        \node [block-adv] (a3) at (2,-3) {3};
        \node [block-adv] (a4) at (2,-4) {4};
        \node [block-adv] (a5) at (3,-4) {5};
        \node [block-adv] (a6) at (4,-4) {6};
        
        \draw [link-adv] (a5.north) -- (a3);
        \draw [link-adv] (a6.north) -- (a3);
        \draw [link-adv] (a4.north) -- (a3);
        \draw [link-adv] (a3.north) -- (a2);
        
        \node [block-hon] (h11) at (2,-5) {11};
        \node [block-hon] (h12) at (2,-6) {12};
        
        \draw [link-hon] (h12) -- (h11);
        \draw [link-hon] (h11) -- (a4);
        
        \node [block-adv] (a5) at (3,-5) {5};
        \node [block-adv] (a6) at (3,-6) {6};
        
        \draw [link-adv] (a6.north) -- (a5);
        \draw [link-adv] (a5.north) -- (a4);
        
        \node [block-hon,opacity=0.3] (hNEW) at (3,-7) {...};
        \draw [link-hon,opacity=0.3] (hNEW) -- (a6);
    
    \end{tikzpicture}
    \caption{Finally, note that the adversary can reuse its blocks $5$, $6$. Honest nodes build a new chain on top of $4 \rightarrow 3 \rightarrow 2 \rightarrow 1 \rightarrow \mathsf{Genesis}$. Once the new chain reaches length $2$, the adversary releases the last displacing sub-tree.}
    \label{fig:avalanche-attack-4}
\end{figure}

We illustrate the attack using a slightly simplified example where the adversary starts with $k=6$ withheld blocks and does not gain any new blocks during the attack.
In this case, the attack eventually runs out of steam and stops. 
(In reality, the larger the number of withheld blocks, the more likely the attack continues practically forever, and even for low $k$ that probability is not negligible.) 
Still, the example illustrates that $k=6$ blocks are enough for the adversary to displace $12$ honest blocks---not a good sign.

First, the adversary withholds its flat-but-wide sub-tree of $k=6$ withheld blocks, while honest nodes produce a chain (\cf Figure~\ref{fig:avalanche-attack-1}).
(Green and red indicate honest and adversarial blocks respectively, and the numbers on blocks indicate which block production opportunity of honest/adversary they correspond to.)
Once honest nodes build a chain of length $k=6$, the adversary releases the withheld blocks, and displaces the honest chain (\cf Figure~\ref{fig:avalanche-attack-2}).
Note that the adversary can reuse its blocks $3, 4, 5, 6$. 
Honest nodes build a new chain on top of $2 \rightarrow 1 \rightarrow \mathsf{Genesis}$. 
Once that new chain reaches length $4$, the adversary releases another displacing sub-tree (\cf Figure~\ref{fig:avalanche-attack-3}).
Finally, note that the adversary can again reuse its blocks $5$, $6$.
Honest nodes build a new chain on top of $4 \rightarrow 3 \rightarrow 2 \rightarrow 1 \rightarrow \mathsf{Genesis}$. 
Once the new chain reaches length $2$, the adversary releases the last displacing sub-tree (\cf Figure~\ref{fig:avalanche-attack-4}).
Honest nodes now build on $6 \rightarrow 5 \rightarrow 4 \rightarrow 3 \rightarrow 2 \rightarrow 1 \rightarrow \mathsf{Genesis}$. 
All honest blocks so far have been displaced. 
Overall, with this strategy, the adversary gets to displace $\Theta(k^2)$ honest blocks with $k$ withheld adversarial blocks.

\subsection{Attack Details}
\label{sec:avalanche-attack-details}

Selfish mining and equivocations can be used to attack PoS GHOST (using an `avalanche of equivocating sub-trees rolling over honest chains'---hence the name of the attack). 
The following description is for vanilla PoS GHOST, but can be straightforwardly translated for Committee-GHOST. 
Variants of this attack work for Committee-GHOST with proposer boosting \cite{mitigationlmdghostbalancingattacks} as well.

Suppose an adversary gets $k$ block production opportunities in a row, for modest $k$. This eventually happens with considerable probability.
The adversary withholds these $k$ blocks, as in \emph{selfish mining} (\cf Figure~\ref{fig:avalanche-attack-1} above). 
On average, more honest blocks are produced than adversary blocks, so the developing honest chain eventually `catches up' with the $k$ withheld adversarial blocks.
In that moment, the adversary releases the $k$ withheld blocks. 
However, not on a competing adversarial chain (as in selfish mining for a Longest Chain protocol), but on a competing adversarial sub-tree of height $2$, where all but the first withheld block are siblings, and children of the first withheld block. 
Due to the GHOST weight counting, this adversarial sub-tree is now of equal weight as the honest chain---so the honest chain is abandoned (\cf Figure~\ref{fig:avalanche-attack-2} above).
At the same time, ties between equal-weight sub-trees are broken such that honest nodes from now on build on what was the second withheld block.
This allows the adversary to reuse in the form of \emph{equivocations} the withheld blocks $3$, $4$, ..., $k$ on top of the chain $2 \rightarrow 1 \rightarrow \mathsf{Genesis}$ formed by the first two withheld adversarial blocks, which is now the chain adopted by the honest nodes.

As an overall result of the attack so far, the adversary started with $k$ withheld blocks, has used those to displace $k$ honest blocks, and is now left with equivocating copies of $k-2$ adversarial withheld blocks that it can still reuse through equivocations (\cf Figure~\ref{fig:avalanche-attack-3} above). 
In addition, while the $k$ honest blocks were produced, the adversary likely had a few block production opportunities of its own, which get added to the pool of adversarial withheld blocks. 
Note that the attack has renewed in favor of the adversary if the adversary had two new block production opportunities, making up for the two adversarial withheld blocks lost because they cannot be reused.
The process is now repeated (\cf Figure~\ref{fig:avalanche-attack-4} above): 
The adversary has a bunch withheld blocks; whenever honest nodes have built a chain of weight equal to the withheld blocks, then the adversary releases a competing sub-tree of height $2$; 
the chain made up from the first two released withheld blocks is adopted by the honest nodes, the other block production opportunities can still be reused in the future through equivocations on top of it and thus remain in the pool of withheld blocks of the adversary.

If the adversary starts out with enough withheld blocks $k$, and adversarial stake is not too small, then the adversary gains $2$ block production opportunities during the production of the $k$ honest blocks that will be displaced subsequently, and the process renews (or even drifts in favor of the adversary). 
No honest blocks enter the canonical chain permanently---a liveness failure.
Also, all honest blocks are eventually removed from the canonical chain, after increasing amount of time---a safety failure for any fixed confirmation time.

\subsection{Proof-of-Concept Implementation}
\label{sec:avalanche-implementation}

\begin{figure}
    \centering
    \includegraphics[width=\linewidth]{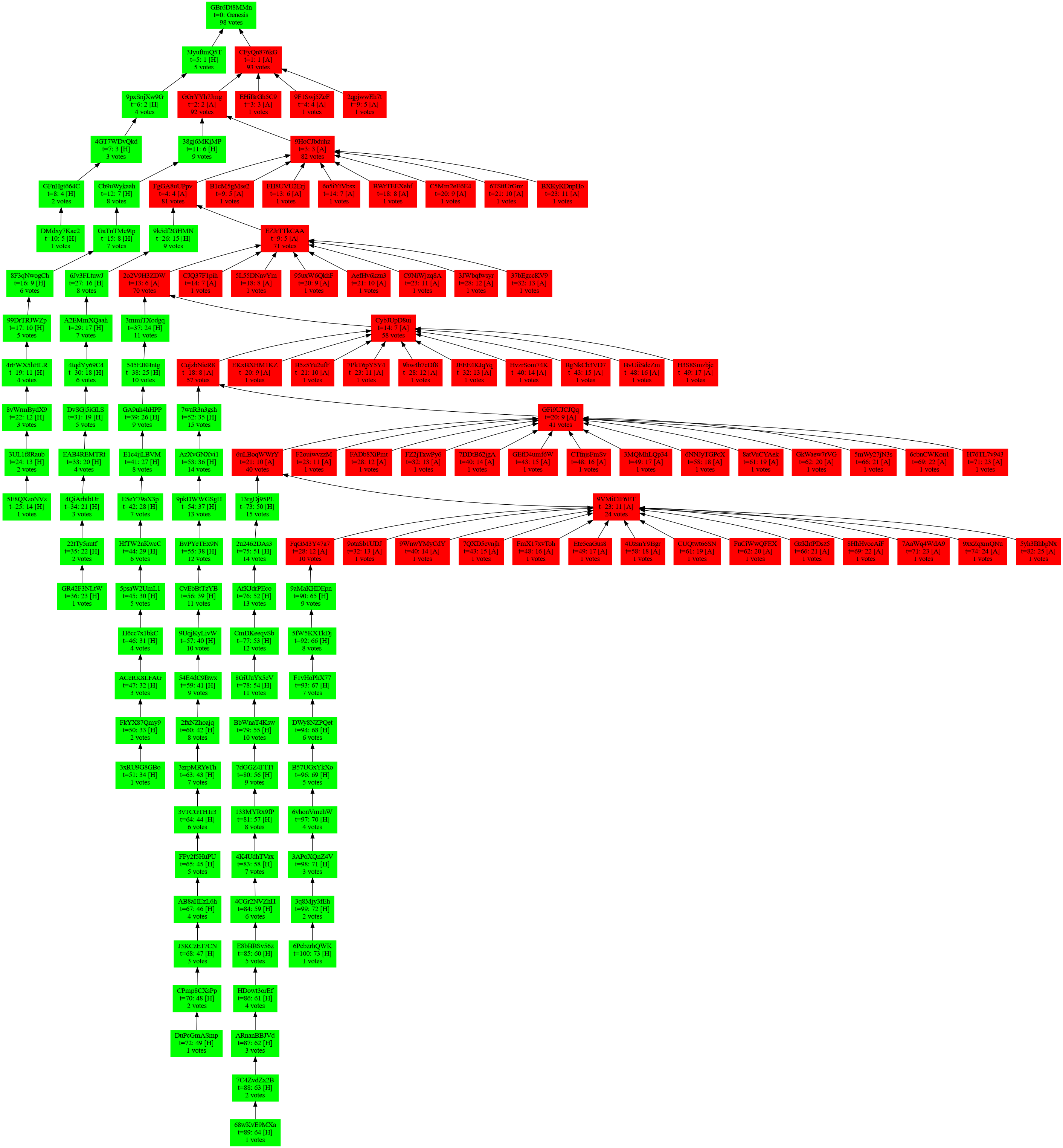}
    \caption{Snapshot of the block tree resulting after $100$ time slots in our proof-of-concept implementation of the avalanche attack on PoS GHOST (adversarial blocks: red, honest blocks: green; adversarial stake: $30\%$, initially withheld adversarial blocks: $4$)}
    \label{fig:avalanche-attack-pos-ghost}
\end{figure}

\begin{figure}
    \centering
    \includegraphics[width=\linewidth]{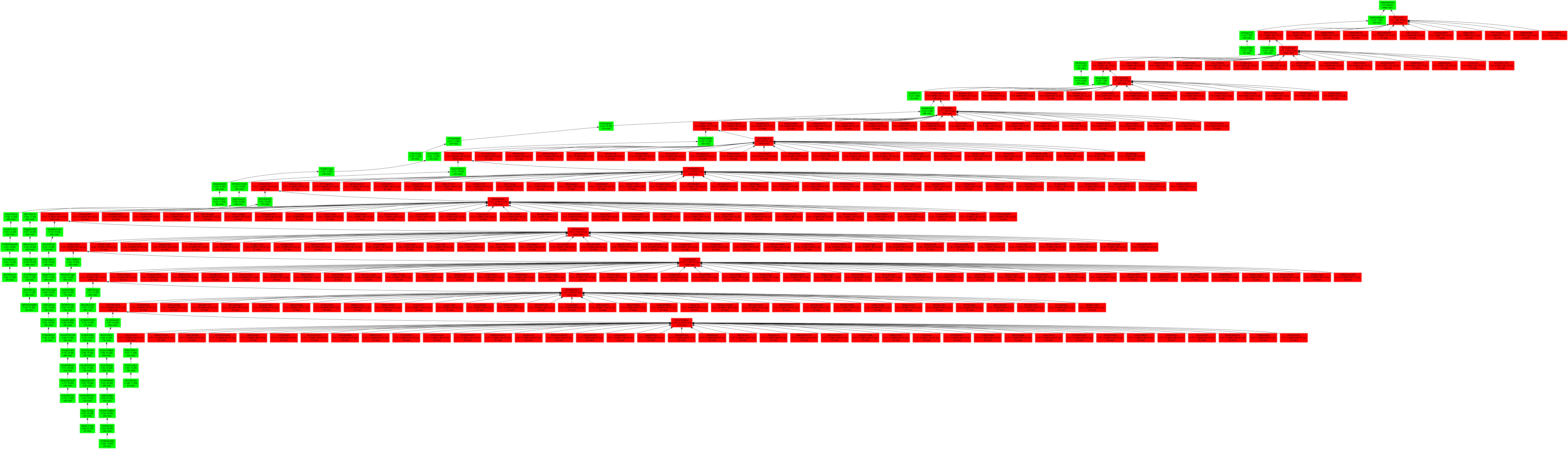}
    \caption{Snapshot of the block tree resulting after $100$ time slots in our proof-of-concept implementation of the avalanche attack on Committee-GHOST (adversarial blocks: red, honest blocks: green; adversarial stake: $20\%$, initially withheld adversarial blocks: $12$)}
    \label{fig:avalanche-attack-committee-ghost}
\end{figure}

For illustration purposes, we plot a snapshot of the block tree resulting after $100$ time slots in our proof-of-concept implementation---see Figure~\ref{fig:avalanche-attack-pos-ghost} for PoS GHOST and Figure~\ref{fig:avalanche-attack-committee-ghost} for Committee-GHOST.\footnote{Source code: \url{https://github.com/tse-group/pos-ghost-attack}} 
The attack is still ongoing thereafter, and as long as the attack is sustained, no honest blocks remain in the canonical chain permanently.

\subsection{Applicability to PoS Ethereum}
\label{sec:avalanche-applicability}

Section~\ref{sec:modelprotocol-protocol} gives a description of the Committee-GHOST protocol based on the consensus protocol of PoS Ethereum, albeit without the LMD aspect.
The avalanche attack works on this protocol as well.
In particular, an adversary controlling any positive fraction of validators can still replace $\Theta(k^2)$ honest blocks with $k$ withheld blocks in this protocol.

PoS Ethereum's LMD (Latest Message Driven) aspect interferes with this attack, but comes with its own challenges, as shown in Section~\ref{sec:attack-lmd-ghost}.
Nevertheless, the avalanche attack suggests a fundamental conceptual incompatibility between PoS (which enables equivocations) and GHOST (where equivocating uncle blocks disproportionately influence fork choice) and casts doubt over whether GHOST should be used with PoS at all.

\section{LMD-Specific Balancing Attack on PoS Ethereum}
\label{sec:attack-lmd-ghost}

Proposal weights (also called `proposer boosting') were suggested \cite{mitigationlmdghostbalancingattacks} and implemented\footnote{\url{https://github.com/ethereum/consensus-specs/pull/2730}} 
to mitigate earlier balancing attacks \cite{ethresearch-balancing-attack,ethresearch-balancing-attack2}.
However, we show that the LMD aspect of PoS Ethereum's fork choice enables balancing attacks even with proposal weights. 
This is particularly dire because PoS GHOST without LMD is susceptible to the avalanche attack, as described in Section~\ref{sec:avalanche-attack}.
A version of PoS Ethereum with the LMD rule is described in Section~\ref{sec:modelprotocol-protocol-lmd}.

\subsection{Preliminaries}
\label{sec:attack-lmd-ghost-prelim}

Recall the following from earlier discussion of balancing-type attacks \cite{ethresearch-balancing-attack,ethresearch-balancing-attack2}:
\begin{itemize}
    \item On a high level, the balancing attack consists of two steps: First, adversarial block proposers initiate two competing chains---let us call them $\mathsf{Left}$ and $\mathsf{Right}$. Then, a handful of adversarial votes per slot, released under particular circumstances, suffice to steer honest validators’ votes so as to keep the system in a tie between the two chains and consequently stall consensus.
    
    \item It is quite feasible for an adversary to release two messages to the network in such a way that roughly half of the honest validators receive one message first and the other half of the honest validators receives the other message first. This certainly holds in the setting of adversarial network propagation delay \cite{ethresearch-balancing-attack} but also in the weaker setting of random network propagation delay \cite{ethresearch-balancing-attack2}.
    
    \item The LMD rule deals with equivocating votes in the following way. Under LMD, every validator keeps a table of the `latest message' (here, each message is a vote) received from each other validator, in the following manner:\footnote{\url{https://github.com/ethereum/consensus-specs/blob/72d45971310a24f6e5ecfb149d23c9fac4c7878a/specs/phase0/fork-choice.md\#update_latest_messages}} When a valid vote from a validator is received, then the `latest message' table entry for that validator is updated if and only if the new vote is from a time slot \emph{strictly later than} the current 'latest message' table entry. Thus, if a validator observes two equivocating votes from the same validator for the \emph{same} time slot, the validator considers the vote received earlier in time.
\end{itemize}

\subsection{High Level Description}
\label{sec:attack-lmd-ghost-high-level}

The LMD rule gives the adversary a remarkable power in a balancing attack: 
Once the adversary has set up two competing chains, it can equivocate on them. 
The release of these equivocating votes can be timed such that the vote for $\mathsf{Left}$ is received by half of honest validators first, and the vote for $\mathsf{Right}$ is received by the other half of honest validators first. 
Honest validators are split in their views concerning the `latest messages' from adversarial validators. 
Even though all validators will soon have received both votes, the split view persists for a considerable time due to the LMD rule, and since the adversarial validators release no votes for later slots.

As a result, half of the honest validators will see $\mathsf{Left}$ as leading, and will vote for it; half will see $\mathsf{Right}$ as leading, and will vote for it. 
But since the honest validators are split roughly in half, their votes balance, and they continue to see their respective chain as leading. 
(The adversary might have to release a few votes every now and then to counteract any drift stemming from an imbalance on the chains different honest validators see as leading.) 
This effect is so stark, that it could only be overcome using proposer boosting if the proposal weight exceeds the adversary's equivocating votes (which is some fraction of the committee size) by more than a constant factor. 
Otherwise, if the adversary leads that constant factor number of slots, it can surpass the proposer boost again.
In that case, the proposer effectively overpowers the committees by far, thus eliminating the purpose of committees.

\subsection{A Simple Example}
\label{sec:attack-lmd-ghost-simple-example}

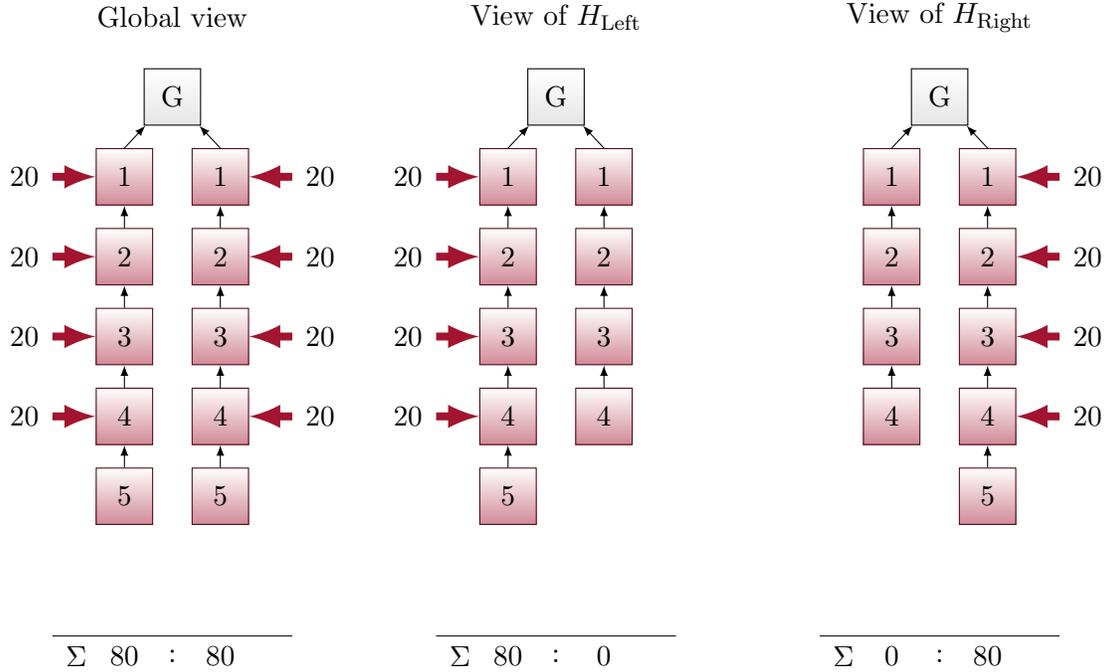
\begin{figure}
    \centering
    \begin{tikzpicture}[blockchain,x=0.75cm,scale=0.85]
    
        \begin{scope}
    
            \node at (0,1) {Global view};
            
            \node at (-1,-7) {$80$};
            \node at (1,-7) {$80$};
            \node at (-2,-7) {$\Sigma$};
            \node at (0,-7) {:};
            \draw (-2.5,-6.75) -- (2.5,-6.75);
    
            \node [block] (G) at (0,0) {G};
            
            \node [block-adv] (a11) at (-1,-1) {1};
            \node [block-adv] (a12) at (+1,-1) {1};
            \node [block-adv] (a21) at (-1,-2) {2};
            \node [block-adv] (a22) at (+1,-2) {2};
            \node [block-adv] (a31) at (-1,-3) {3};
            \node [block-adv] (a32) at (+1,-3) {3};
            \node [block-adv] (a41) at (-1,-4) {4};
            \node [block-adv] (a42) at (+1,-4) {4};
            \node [block-adv] (a51) at (-1,-5) {5};
            \node [block-adv] (a52) at (+1,-5) {5};
            
            \draw [link-adv] (a11.north) -- (G);
            \draw [link-adv] (a12.north) -- (G);
            \draw [link-adv] (a21.north) -- (a11);
            \draw [link-adv] (a22.north) -- (a12);
            \draw [link-adv] (a31.north) -- (a21);
            \draw [link-adv] (a32.north) -- (a22);
            \draw [link-adv] (a41.north) -- (a31);
            \draw [link-adv] (a42.north) -- (a32);
            \draw [link-adv] (a51.north) -- (a41);
            \draw [link-adv] (a52.north) -- (a42);

            \draw [latex-,line width=3pt,draw=myParula07Red] (a11) -- ++(-1.5,0) node [left] {$20$};
            \draw [latex-,line width=3pt,draw=myParula07Red] (a21) -- ++(-1.5,0) node [left] {$20$};
            \draw [latex-,line width=3pt,draw=myParula07Red] (a31) -- ++(-1.5,0) node [left] {$20$};
            \draw [latex-,line width=3pt,draw=myParula07Red] (a41) -- ++(-1.5,0) node [left] {$20$};
            \draw [latex-,line width=3pt,draw=myParula07Red] (a12) -- ++(1.5,0) node [right] {$20$};
            \draw [latex-,line width=3pt,draw=myParula07Red] (a22) -- ++(1.5,0) node [right] {$20$};
            \draw [latex-,line width=3pt,draw=myParula07Red] (a32) -- ++(1.5,0) node [right] {$20$};
            \draw [latex-,line width=3pt,draw=myParula07Red] (a42) -- ++(1.5,0) node [right] {$20$};

        \end{scope}
    
        \begin{scope}[xshift=6cm]
    
            \node at (0,1) {View of $H_{\mathrm{Left}}$};
            
            \node at (-1,-7) {$80$};
            \node at (1,-7) {$0$};
            \node at (-2,-7) {$\Sigma$};
            \node at (0,-7) {:};
            \draw (-2.5,-6.75) -- (2.5,-6.75);
    
            \node [block] (G) at (0,0) {G};
            
            \node [block-adv] (a11) at (-1,-1) {1};
            \node [block-adv] (a12) at (+1,-1) {1};
            \node [block-adv] (a21) at (-1,-2) {2};
            \node [block-adv] (a22) at (+1,-2) {2};
            \node [block-adv] (a31) at (-1,-3) {3};
            \node [block-adv] (a32) at (+1,-3) {3};
            \node [block-adv] (a41) at (-1,-4) {4};
            \node [block-adv] (a42) at (+1,-4) {4};
            \node [block-adv] (a51) at (-1,-5) {5};

            \draw [link-adv] (a11.north) -- (G);
            \draw [link-adv] (a12.north) -- (G);
            \draw [link-adv] (a21.north) -- (a11);
            \draw [link-adv] (a22.north) -- (a12);
            \draw [link-adv] (a31.north) -- (a21);
            \draw [link-adv] (a32.north) -- (a22);
            \draw [link-adv] (a41.north) -- (a31);
            \draw [link-adv] (a42.north) -- (a32);
            \draw [link-adv] (a51.north) -- (a41);

            \draw [latex-,line width=3pt,draw=myParula07Red] (a11) -- ++(-1.5,0) node [left] {$20$};
            \draw [latex-,line width=3pt,draw=myParula07Red] (a21) -- ++(-1.5,0) node [left] {$20$};
            \draw [latex-,line width=3pt,draw=myParula07Red] (a31) -- ++(-1.5,0) node [left] {$20$};
            \draw [latex-,line width=3pt,draw=myParula07Red] (a41) -- ++(-1.5,0) node [left] {$20$};

        \end{scope}
    
        \begin{scope}[xshift=12cm]
    
            \node at (0,1) {View of $H_{\mathrm{Right}}$};
            
            \node at (-1,-7) {$0$};
            \node at (1,-7) {$80$};
            \node at (-2,-7) {$\Sigma$};
            \node at (0,-7) {:};
            \draw (-2.5,-6.75) -- (2.5,-6.75);
    
            \node [block] (G) at (0,0) {G};
            
            \node [block-adv] (a11) at (-1,-1) {1};
            \node [block-adv] (a12) at (+1,-1) {1};
            \node [block-adv] (a21) at (-1,-2) {2};
            \node [block-adv] (a22) at (+1,-2) {2};
            \node [block-adv] (a31) at (-1,-3) {3};
            \node [block-adv] (a32) at (+1,-3) {3};
            \node [block-adv] (a41) at (-1,-4) {4};
            \node [block-adv] (a42) at (+1,-4) {4};
            \node [block-adv] (a52) at (+1,-5) {5};
            
            \draw [link-adv] (a11.north) -- (G);
            \draw [link-adv] (a12.north) -- (G);
            \draw [link-adv] (a21.north) -- (a11);
            \draw [link-adv] (a22.north) -- (a12);
            \draw [link-adv] (a31.north) -- (a21);
            \draw [link-adv] (a32.north) -- (a22);
            \draw [link-adv] (a41.north) -- (a31);
            \draw [link-adv] (a42.north) -- (a32);
            \draw [link-adv] (a52.north) -- (a42);

            \draw [latex-,line width=3pt,draw=myParula07Red] (a12) -- ++(1.5,0) node [right] {$20$};
            \draw [latex-,line width=3pt,draw=myParula07Red] (a22) -- ++(1.5,0) node [right] {$20$};
            \draw [latex-,line width=3pt,draw=myParula07Red] (a32) -- ++(1.5,0) node [right] {$20$};
            \draw [latex-,line width=3pt,draw=myParula07Red] (a42) -- ++(1.5,0) node [right] {$20$};

        \end{scope}
    
    \end{tikzpicture}
    \caption{By the LMD rule, validators in $H_{\mathrm{Left}}$ and $H_{\mathrm{Right}}$ believe that $\mathsf{Left}$ and $\mathsf{Right}$ have $80$ votes, respectively. They also believe that the respective other chain has $0$ votes (as the later arriving votes are not considered due to LMD).}
    \label{fig:lmd-attack-1}
\end{figure}

\begin{figure}
    \centering
    \begin{tikzpicture}[blockchain,x=0.75cm,scale=0.85]
    
        \begin{scope}
    
            \node at (0,1) {Global view};
            
            \node at (-1,-7) {$80$};
            \node at (1,-7) {$80$};
            \node at (-2,-7) {$\Sigma$};
            \node at (0,-7) {:};
            \draw (-2.5,-6.75) -- (2.5,-6.75);
    
            \node [block] (G) at (0,0) {G};
            
            \node [block-adv] (a11) at (-1,-1) {1};
            \node [block-adv] (a12) at (+1,-1) {1};
            \node [block-adv] (a21) at (-1,-2) {2};
            \node [block-adv] (a22) at (+1,-2) {2};
            \node [block-adv] (a31) at (-1,-3) {3};
            \node [block-adv] (a32) at (+1,-3) {3};
            \node [block-adv] (a41) at (-1,-4) {4};
            \node [block-adv] (a42) at (+1,-4) {4};
            \node [block-adv] (a51) at (-1,-5) {5};
            \node [block-adv] (a52) at (+1,-5) {5};
            
            \draw [link-adv] (a11.north) -- (G);
            \draw [link-adv] (a12.north) -- (G);
            \draw [link-adv] (a21.north) -- (a11);
            \draw [link-adv] (a22.north) -- (a12);
            \draw [link-adv] (a31.north) -- (a21);
            \draw [link-adv] (a32.north) -- (a22);
            \draw [link-adv] (a41.north) -- (a31);
            \draw [link-adv] (a42.north) -- (a32);
            \draw [link-adv] (a51.north) -- (a41);
            \draw [link-adv] (a52.north) -- (a42);

            \draw [latex-,line width=3pt,draw=myParula07Red] (a11) -- ++(-1.5,0) node [left] {$20$};
            \draw [latex-,line width=3pt,draw=myParula07Red] (a21) -- ++(-1.5,0) node [left] {$20$};
            \draw [latex-,line width=3pt,draw=myParula07Red] (a31) -- ++(-1.5,0) node [left] {$20$};
            \draw [latex-,line width=3pt,draw=myParula07Red] (a41) -- ++(-1.5,0) node [left] {$20$};
            \draw [latex-,line width=3pt,draw=myParula07Red] (a12) -- ++(1.5,0) node [right] {$20$};
            \draw [latex-,line width=3pt,draw=myParula07Red] (a22) -- ++(1.5,0) node [right] {$20$};
            \draw [latex-,line width=3pt,draw=myParula07Red] (a32) -- ++(1.5,0) node [right] {$20$};
            \draw [latex-,line width=3pt,draw=myParula07Red] (a42) -- ++(1.5,0) node [right] {$20$};

        \end{scope}
    
        \begin{scope}[xshift=6cm]
    
            \node at (0,1) {View of $H_{\mathrm{Left}}$};
            
            \node at (-1,-7) {$80$};
            \node at (1,-7) {$0$};
            \node at (-2,-7) {$\Sigma$};
            \node at (0,-7) {:};
            \draw (-2.5,-6.75) -- (2.5,-6.75);
    
            \node [block] (G) at (0,0) {G};
            
            \node [block-adv] (a11) at (-1,-1) {1};
            \node [block-adv] (a12) at (+1,-1) {1};
            \node [block-adv] (a21) at (-1,-2) {2};
            \node [block-adv] (a22) at (+1,-2) {2};
            \node [block-adv] (a31) at (-1,-3) {3};
            \node [block-adv] (a32) at (+1,-3) {3};
            \node [block-adv] (a41) at (-1,-4) {4};
            \node [block-adv] (a42) at (+1,-4) {4};
            \node [block-adv] (a51) at (-1,-5) {5};
            \node [block-adv] (a52) at (+1,-5) {5};
            
            \draw [link-adv] (a11.north) -- (G);
            \draw [link-adv] (a12.north) -- (G);
            \draw [link-adv] (a21.north) -- (a11);
            \draw [link-adv] (a22.north) -- (a12);
            \draw [link-adv] (a31.north) -- (a21);
            \draw [link-adv] (a32.north) -- (a22);
            \draw [link-adv] (a41.north) -- (a31);
            \draw [link-adv] (a42.north) -- (a32);
            \draw [link-adv] (a51.north) -- (a41);
            \draw [link-adv] (a52.north) -- (a42);

            \draw [latex-,line width=3pt,draw=myParula07Red] (a11) -- ++(-1.5,0) node [left] {$20$};
            \draw [latex-,line width=3pt,draw=myParula07Red] (a21) -- ++(-1.5,0) node [left] {$20$};
            \draw [latex-,line width=3pt,draw=myParula07Red] (a31) -- ++(-1.5,0) node [left] {$20$};
            \draw [latex-,line width=3pt,draw=myParula07Red] (a41) -- ++(-1.5,0) node [left] {$20$};
            \draw [latex-,line width=3pt,draw=myParula07Red,opacity=0.3] (a12) -- ++(1.5,0) node [right] {$20$};
            \draw [latex-,line width=3pt,draw=myParula07Red,opacity=0.3] (a22) -- ++(1.5,0) node [right] {$20$};
            \draw [latex-,line width=3pt,draw=myParula07Red,opacity=0.3] (a32) -- ++(1.5,0) node [right] {$20$};
            \draw [latex-,line width=3pt,draw=myParula07Red,opacity=0.3] (a42) -- ++(1.5,0) node [right] {$20$};

        \end{scope}
    
        \begin{scope}[xshift=12cm]
    
            \node at (0,1) {View of $H_{\mathrm{Right}}$};
            
            \node at (-1,-7) {$0$};
            \node at (1,-7) {$80$};
            \node at (-2,-7) {$\Sigma$};
            \node at (0,-7) {:};
            \draw (-2.5,-6.75) -- (2.5,-6.75);
    
            \node [block] (G) at (0,0) {G};
            
            \node [block-adv] (a11) at (-1,-1) {1};
            \node [block-adv] (a12) at (+1,-1) {1};
            \node [block-adv] (a21) at (-1,-2) {2};
            \node [block-adv] (a22) at (+1,-2) {2};
            \node [block-adv] (a31) at (-1,-3) {3};
            \node [block-adv] (a32) at (+1,-3) {3};
            \node [block-adv] (a41) at (-1,-4) {4};
            \node [block-adv] (a42) at (+1,-4) {4};
            \node [block-adv] (a51) at (-1,-5) {5};
            \node [block-adv] (a52) at (+1,-5) {5};
            
            \draw [link-adv] (a11.north) -- (G);
            \draw [link-adv] (a12.north) -- (G);
            \draw [link-adv] (a21.north) -- (a11);
            \draw [link-adv] (a22.north) -- (a12);
            \draw [link-adv] (a31.north) -- (a21);
            \draw [link-adv] (a32.north) -- (a22);
            \draw [link-adv] (a41.north) -- (a31);
            \draw [link-adv] (a42.north) -- (a32);
            \draw [link-adv] (a51.north) -- (a41);
            \draw [link-adv] (a52.north) -- (a42);

            \draw [latex-,line width=3pt,draw=myParula07Red,opacity=0.3] (a11) -- ++(-1.5,0) node [left] {$20$};
            \draw [latex-,line width=3pt,draw=myParula07Red,opacity=0.3] (a21) -- ++(-1.5,0) node [left] {$20$};
            \draw [latex-,line width=3pt,draw=myParula07Red,opacity=0.3] (a31) -- ++(-1.5,0) node [left] {$20$};
            \draw [latex-,line width=3pt,draw=myParula07Red,opacity=0.3] (a41) -- ++(-1.5,0) node [left] {$20$};
            \draw [latex-,line width=3pt,draw=myParula07Red] (a12) -- ++(1.5,0) node [right] {$20$};
            \draw [latex-,line width=3pt,draw=myParula07Red] (a22) -- ++(1.5,0) node [right] {$20$};
            \draw [latex-,line width=3pt,draw=myParula07Red] (a32) -- ++(1.5,0) node [right] {$20$};
            \draw [latex-,line width=3pt,draw=myParula07Red] (a42) -- ++(1.5,0) node [right] {$20$};

        \end{scope}
    
    \end{tikzpicture}
    \caption{Suppose the validator of slot $6$ is honest and from set $H_{\mathrm{Left}}$. Then, it proposes a block extending $\mathsf{Left}$, which gains a proposal boost equivalent to $70$ votes.}
    \label{fig:lmd-attack-2}
\end{figure}
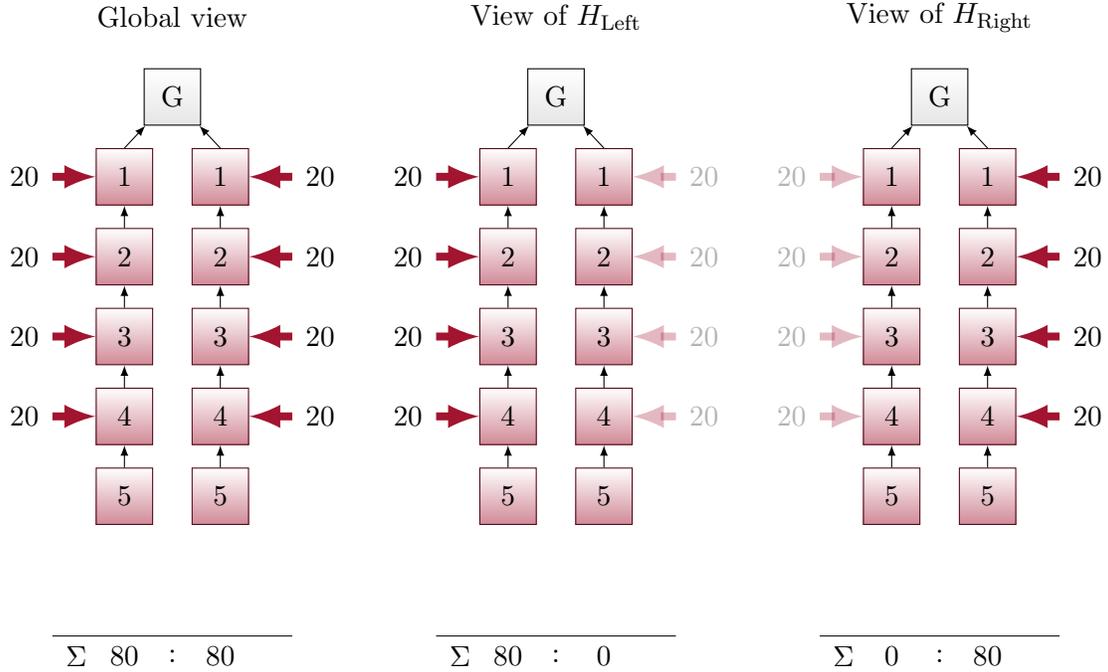

\begin{figure}
    \centering
    \begin{tikzpicture}[blockchain,x=0.75cm,scale=0.85]
    
        \begin{scope}
    
            \node at (0,1) {Global view};
            
            \node at (-1,-7) {$150$};
            \node at (1,-7) {$80$};
            \node at (-2,-7) {$\Sigma$};
            \node at (0,-7) {:};
            \draw (-2.5,-6.75) -- (2.5,-6.75);
    
            \node [block] (G) at (0,0) {G};
            
            \node [block-adv] (a11) at (-1,-1) {1};
            \node [block-adv] (a12) at (+1,-1) {1};
            \node [block-adv] (a21) at (-1,-2) {2};
            \node [block-adv] (a22) at (+1,-2) {2};
            \node [block-adv] (a31) at (-1,-3) {3};
            \node [block-adv] (a32) at (+1,-3) {3};
            \node [block-adv] (a41) at (-1,-4) {4};
            \node [block-adv] (a42) at (+1,-4) {4};
            \node [block-adv] (a51) at (-1,-5) {5};
            \node [block-adv] (a52) at (+1,-5) {5};
            
            \draw [link-adv] (a11.north) -- (G);
            \draw [link-adv] (a12.north) -- (G);
            \draw [link-adv] (a21.north) -- (a11);
            \draw [link-adv] (a22.north) -- (a12);
            \draw [link-adv] (a31.north) -- (a21);
            \draw [link-adv] (a32.north) -- (a22);
            \draw [link-adv] (a41.north) -- (a31);
            \draw [link-adv] (a42.north) -- (a32);
            \draw [link-adv] (a51.north) -- (a41);
            \draw [link-adv] (a52.north) -- (a42);
            
            \draw [latex-,line width=3pt,draw=myParula07Red] (a11) -- ++(-1.5,0) node [left] {$20$};
            \draw [latex-,line width=3pt,draw=myParula07Red] (a21) -- ++(-1.5,0) node [left] {$20$};
            \draw [latex-,line width=3pt,draw=myParula07Red] (a31) -- ++(-1.5,0) node [left] {$20$};
            \draw [latex-,line width=3pt,draw=myParula07Red] (a41) -- ++(-1.5,0) node [left] {$20$};
            \draw [latex-,line width=3pt,draw=myParula07Red] (a12) -- ++(1.5,0) node [right] {$20$};
            \draw [latex-,line width=3pt,draw=myParula07Red] (a22) -- ++(1.5,0) node [right] {$20$};
            \draw [latex-,line width=3pt,draw=myParula07Red] (a32) -- ++(1.5,0) node [right] {$20$};
            \draw [latex-,line width=3pt,draw=myParula07Red] (a42) -- ++(1.5,0) node [right] {$20$};

            \node [block-hon] (h1) at (-1,-6) {6};
            \draw [link-hon] (h1.north) -- (a51);
            \draw [latex-,line width=3pt,draw=black!50] (h1) -- ++(1.5,0) node [right] {$70$};
            
        \end{scope}
    
        \begin{scope}[xshift=6cm]
    
            \node at (0,1) {View of $H_{\mathrm{Left}}$};
            
            \node at (-1,-7) {$150$};
            \node at (1,-7) {$0$};
            \node at (-2,-7) {$\Sigma$};
            \node at (0,-7) {:};
            \draw (-2.5,-6.75) -- (2.5,-6.75);
    
            \node [block] (G) at (0,0) {G};
            
            \node [block-adv] (a11) at (-1,-1) {1};
            \node [block-adv] (a12) at (+1,-1) {1};
            \node [block-adv] (a21) at (-1,-2) {2};
            \node [block-adv] (a22) at (+1,-2) {2};
            \node [block-adv] (a31) at (-1,-3) {3};
            \node [block-adv] (a32) at (+1,-3) {3};
            \node [block-adv] (a41) at (-1,-4) {4};
            \node [block-adv] (a42) at (+1,-4) {4};
            \node [block-adv] (a51) at (-1,-5) {5};
            \node [block-adv] (a52) at (+1,-5) {5};
            
            \draw [link-adv] (a11.north) -- (G);
            \draw [link-adv] (a12.north) -- (G);
            \draw [link-adv] (a21.north) -- (a11);
            \draw [link-adv] (a22.north) -- (a12);
            \draw [link-adv] (a31.north) -- (a21);
            \draw [link-adv] (a32.north) -- (a22);
            \draw [link-adv] (a41.north) -- (a31);
            \draw [link-adv] (a42.north) -- (a32);
            \draw [link-adv] (a51.north) -- (a41);
            \draw [link-adv] (a52.north) -- (a42);

            \draw [latex-,line width=3pt,draw=myParula07Red] (a11) -- ++(-1.5,0) node [left] {$20$};
            \draw [latex-,line width=3pt,draw=myParula07Red] (a21) -- ++(-1.5,0) node [left] {$20$};
            \draw [latex-,line width=3pt,draw=myParula07Red] (a31) -- ++(-1.5,0) node [left] {$20$};
            \draw [latex-,line width=3pt,draw=myParula07Red] (a41) -- ++(-1.5,0) node [left] {$20$};
            \draw [latex-,line width=3pt,draw=myParula07Red,opacity=0.3] (a12) -- ++(1.5,0) node [right] {$20$};
            \draw [latex-,line width=3pt,draw=myParula07Red,opacity=0.3] (a22) -- ++(1.5,0) node [right] {$20$};
            \draw [latex-,line width=3pt,draw=myParula07Red,opacity=0.3] (a32) -- ++(1.5,0) node [right] {$20$};
            \draw [latex-,line width=3pt,draw=myParula07Red,opacity=0.3] (a42) -- ++(1.5,0) node [right] {$20$};

            \node [block-hon] (h1) at (-1,-6) {6};
            \draw [link-hon] (h1.north) -- (a51);
            \draw [latex-,line width=3pt,draw=black!50] (h1) -- ++(1.5,0) node [right] {$70$};
            
        \end{scope}
    
        \begin{scope}[xshift=12cm]
    
            \node at (0,1) {View of $H_{\mathrm{Right}}$};
            
            \node at (-1,-7) {$70$};
            \node at (1,-7) {$80$};
            \node at (-2,-7) {$\Sigma$};
            \node at (0,-7) {:};
            \draw (-2.5,-6.75) -- (2.5,-6.75);
    
            \node [block] (G) at (0,0) {G};
            
            \node [block-adv] (a11) at (-1,-1) {1};
            \node [block-adv] (a12) at (+1,-1) {1};
            \node [block-adv] (a21) at (-1,-2) {2};
            \node [block-adv] (a22) at (+1,-2) {2};
            \node [block-adv] (a31) at (-1,-3) {3};
            \node [block-adv] (a32) at (+1,-3) {3};
            \node [block-adv] (a41) at (-1,-4) {4};
            \node [block-adv] (a42) at (+1,-4) {4};
            \node [block-adv] (a51) at (-1,-5) {5};
            \node [block-adv] (a52) at (+1,-5) {5};
            
            \draw [link-adv] (a11.north) -- (G);
            \draw [link-adv] (a12.north) -- (G);
            \draw [link-adv] (a21.north) -- (a11);
            \draw [link-adv] (a22.north) -- (a12);
            \draw [link-adv] (a31.north) -- (a21);
            \draw [link-adv] (a32.north) -- (a22);
            \draw [link-adv] (a41.north) -- (a31);
            \draw [link-adv] (a42.north) -- (a32);
            \draw [link-adv] (a51.north) -- (a41);
            \draw [link-adv] (a52.north) -- (a42);

            \draw [latex-,line width=3pt,draw=myParula07Red,opacity=0.3] (a11) -- ++(-1.5,0) node [left] {$20$};
            \draw [latex-,line width=3pt,draw=myParula07Red,opacity=0.3] (a21) -- ++(-1.5,0) node [left] {$20$};
            \draw [latex-,line width=3pt,draw=myParula07Red,opacity=0.3] (a31) -- ++(-1.5,0) node [left] {$20$};
            \draw [latex-,line width=3pt,draw=myParula07Red,opacity=0.3] (a41) -- ++(-1.5,0) node [left] {$20$};
            \draw [latex-,line width=3pt,draw=myParula07Red] (a12) -- ++(1.5,0) node [right] {$20$};
            \draw [latex-,line width=3pt,draw=myParula07Red] (a22) -- ++(1.5,0) node [right] {$20$};
            \draw [latex-,line width=3pt,draw=myParula07Red] (a32) -- ++(1.5,0) node [right] {$20$};
            \draw [latex-,line width=3pt,draw=myParula07Red] (a42) -- ++(1.5,0) node [right] {$20$};

            \node [block-hon] (h1) at (-1,-6) {6};
            \draw [link-hon] (h1.north) -- (a51);
            \draw [latex-,line width=3pt,draw=black!50] (h1) -- ++(1.5,0) node [right] {$70$};
            
        \end{scope}
    
    \end{tikzpicture}
    \caption{Due to the proposal boost, validators in $H_{\mathrm{Left}}$ see $\mathsf{Left}$ as leading with $150$ votes and vote for it. Validators in $H_{\mathrm{Right}}$ see $\mathsf{Left}$ has $70$ votes while $\mathsf{Right}$ has $80$ votes, so they vote for $\mathsf{Right}$.}
    \label{fig:lmd-attack-3}
\end{figure}
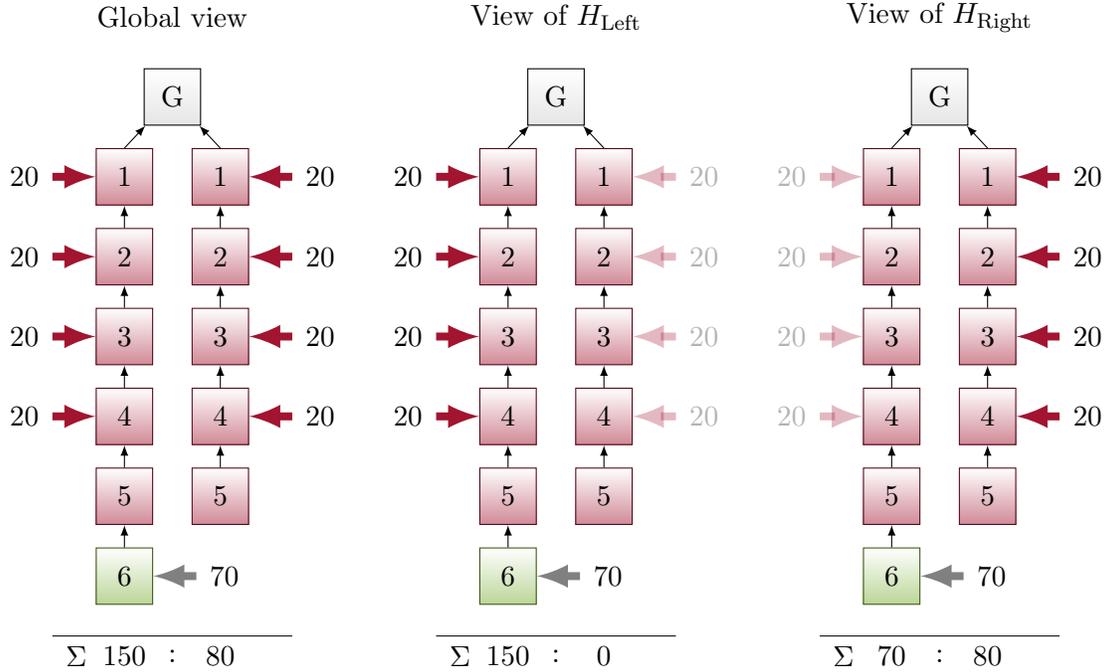

\begin{figure}
    \centering
    \begin{tikzpicture}[blockchain,x=0.75cm,scale=0.85]
    
        \begin{scope}
    
            \node at (0,1) {Global view};
            
            \node at (-1,-7) {$190$};
            \node at (1,-7) {$120$};
            \node at (-2,-7) {$\Sigma$};
            \node at (0,-7) {:};
            \draw (-2.5,-6.75) -- (2.5,-6.75);
    
            \node [block] (G) at (0,0) {G};
            
            \node [block-adv] (a11) at (-1,-1) {1};
            \node [block-adv] (a12) at (+1,-1) {1};
            \node [block-adv] (a21) at (-1,-2) {2};
            \node [block-adv] (a22) at (+1,-2) {2};
            \node [block-adv] (a31) at (-1,-3) {3};
            \node [block-adv] (a32) at (+1,-3) {3};
            \node [block-adv] (a41) at (-1,-4) {4};
            \node [block-adv] (a42) at (+1,-4) {4};
            \node [block-adv] (a51) at (-1,-5) {5};
            \node [block-adv] (a52) at (+1,-5) {5};
            
            \draw [link-adv] (a11.north) -- (G);
            \draw [link-adv] (a12.north) -- (G);
            \draw [link-adv] (a21.north) -- (a11);
            \draw [link-adv] (a22.north) -- (a12);
            \draw [link-adv] (a31.north) -- (a21);
            \draw [link-adv] (a32.north) -- (a22);
            \draw [link-adv] (a41.north) -- (a31);
            \draw [link-adv] (a42.north) -- (a32);
            \draw [link-adv] (a51.north) -- (a41);
            \draw [link-adv] (a52.north) -- (a42);
            
            \draw [latex-,line width=3pt,draw=myParula07Red] (a11) -- ++(-1.5,0) node [left] {$20$};
            \draw [latex-,line width=3pt,draw=myParula07Red] (a21) -- ++(-1.5,0) node [left] {$20$};
            \draw [latex-,line width=3pt,draw=myParula07Red] (a31) -- ++(-1.5,0) node [left] {$20$};
            \draw [latex-,line width=3pt,draw=myParula07Red] (a41) -- ++(-1.5,0) node [left] {$20$};
            \draw [latex-,line width=3pt,draw=myParula07Red] (a12) -- ++(1.5,0) node [right] {$20$};
            \draw [latex-,line width=3pt,draw=myParula07Red] (a22) -- ++(1.5,0) node [right] {$20$};
            \draw [latex-,line width=3pt,draw=myParula07Red] (a32) -- ++(1.5,0) node [right] {$20$};
            \draw [latex-,line width=3pt,draw=myParula07Red] (a42) -- ++(1.5,0) node [right] {$20$};

            \node [block-hon] (h1) at (-1,-6) {6};
            \draw [link-hon] (h1.north) -- (a51);
            \draw [latex-,line width=3pt,draw=black!50] (h1) -- ++(1.5,0) node [right] {$70$};
            
            \draw [latex-,line width=3pt,draw=myParula05Green] (h1) -- ++(-1.5,0) node [left] {$40$};
            \draw [latex-,line width=3pt,draw=myParula05Green] (a52) -- ++(1.5,0) node [right] {$40$};
            
        \end{scope}
    
        \begin{scope}[xshift=6cm]
    
            \node at (0,1) {View of $H_{\mathrm{Left}}$};
            
            \node at (-1,-7) {$190$};
            \node at (1,-7) {$40$};
            \node at (-2,-7) {$\Sigma$};
            \node at (0,-7) {:};
            \draw (-2.5,-6.75) -- (2.5,-6.75);
    
            \node [block] (G) at (0,0) {G};
            
            \node [block-adv] (a11) at (-1,-1) {1};
            \node [block-adv] (a12) at (+1,-1) {1};
            \node [block-adv] (a21) at (-1,-2) {2};
            \node [block-adv] (a22) at (+1,-2) {2};
            \node [block-adv] (a31) at (-1,-3) {3};
            \node [block-adv] (a32) at (+1,-3) {3};
            \node [block-adv] (a41) at (-1,-4) {4};
            \node [block-adv] (a42) at (+1,-4) {4};
            \node [block-adv] (a51) at (-1,-5) {5};
            \node [block-adv] (a52) at (+1,-5) {5};
            
            \draw [link-adv] (a11.north) -- (G);
            \draw [link-adv] (a12.north) -- (G);
            \draw [link-adv] (a21.north) -- (a11);
            \draw [link-adv] (a22.north) -- (a12);
            \draw [link-adv] (a31.north) -- (a21);
            \draw [link-adv] (a32.north) -- (a22);
            \draw [link-adv] (a41.north) -- (a31);
            \draw [link-adv] (a42.north) -- (a32);
            \draw [link-adv] (a51.north) -- (a41);
            \draw [link-adv] (a52.north) -- (a42);

            \draw [latex-,line width=3pt,draw=myParula07Red] (a11) -- ++(-1.5,0) node [left] {$20$};
            \draw [latex-,line width=3pt,draw=myParula07Red] (a21) -- ++(-1.5,0) node [left] {$20$};
            \draw [latex-,line width=3pt,draw=myParula07Red] (a31) -- ++(-1.5,0) node [left] {$20$};
            \draw [latex-,line width=3pt,draw=myParula07Red] (a41) -- ++(-1.5,0) node [left] {$20$};
            \draw [latex-,line width=3pt,draw=myParula07Red,opacity=0.3] (a12) -- ++(1.5,0) node [right] {$20$};
            \draw [latex-,line width=3pt,draw=myParula07Red,opacity=0.3] (a22) -- ++(1.5,0) node [right] {$20$};
            \draw [latex-,line width=3pt,draw=myParula07Red,opacity=0.3] (a32) -- ++(1.5,0) node [right] {$20$};
            \draw [latex-,line width=3pt,draw=myParula07Red,opacity=0.3] (a42) -- ++(1.5,0) node [right] {$20$};

            \node [block-hon] (h1) at (-1,-6) {6};
            \draw [link-hon] (h1.north) -- (a51);
            \draw [latex-,line width=3pt,draw=black!50] (h1) -- ++(1.5,0) node [right] {$70$};
            
            \draw [latex-,line width=3pt,draw=myParula05Green] (h1) -- ++(-1.5,0) node [left] {$40$};
            \draw [latex-,line width=3pt,draw=myParula05Green] (a52) -- ++(1.5,0) node [right] {$40$};
            
        \end{scope}
    
        \begin{scope}[xshift=12cm]
    
            \node at (0,1) {View of $H_{\mathrm{Right}}$};
            
            \node at (-1,-7) {$110$};
            \node at (1,-7) {$120$};
            \node at (-2,-7) {$\Sigma$};
            \node at (0,-7) {:};
            \draw (-2.5,-6.75) -- (2.5,-6.75);
    
            \node [block] (G) at (0,0) {G};
            
            \node [block-adv] (a11) at (-1,-1) {1};
            \node [block-adv] (a12) at (+1,-1) {1};
            \node [block-adv] (a21) at (-1,-2) {2};
            \node [block-adv] (a22) at (+1,-2) {2};
            \node [block-adv] (a31) at (-1,-3) {3};
            \node [block-adv] (a32) at (+1,-3) {3};
            \node [block-adv] (a41) at (-1,-4) {4};
            \node [block-adv] (a42) at (+1,-4) {4};
            \node [block-adv] (a51) at (-1,-5) {5};
            \node [block-adv] (a52) at (+1,-5) {5};
            
            \draw [link-adv] (a11.north) -- (G);
            \draw [link-adv] (a12.north) -- (G);
            \draw [link-adv] (a21.north) -- (a11);
            \draw [link-adv] (a22.north) -- (a12);
            \draw [link-adv] (a31.north) -- (a21);
            \draw [link-adv] (a32.north) -- (a22);
            \draw [link-adv] (a41.north) -- (a31);
            \draw [link-adv] (a42.north) -- (a32);
            \draw [link-adv] (a51.north) -- (a41);
            \draw [link-adv] (a52.north) -- (a42);

            \draw [latex-,line width=3pt,draw=myParula07Red,opacity=0.3] (a11) -- ++(-1.5,0) node [left] {$20$};
            \draw [latex-,line width=3pt,draw=myParula07Red,opacity=0.3] (a21) -- ++(-1.5,0) node [left] {$20$};
            \draw [latex-,line width=3pt,draw=myParula07Red,opacity=0.3] (a31) -- ++(-1.5,0) node [left] {$20$};
            \draw [latex-,line width=3pt,draw=myParula07Red,opacity=0.3] (a41) -- ++(-1.5,0) node [left] {$20$};
            \draw [latex-,line width=3pt,draw=myParula07Red] (a12) -- ++(1.5,0) node [right] {$20$};
            \draw [latex-,line width=3pt,draw=myParula07Red] (a22) -- ++(1.5,0) node [right] {$20$};
            \draw [latex-,line width=3pt,draw=myParula07Red] (a32) -- ++(1.5,0) node [right] {$20$};
            \draw [latex-,line width=3pt,draw=myParula07Red] (a42) -- ++(1.5,0) node [right] {$20$};

            \node [block-hon] (h1) at (-1,-6) {6};
            \draw [link-hon] (h1.north) -- (a51);
            \draw [latex-,line width=3pt,draw=black!50] (h1) -- ++(1.5,0) node [right] {$70$};
            
            \draw [latex-,line width=3pt,draw=myParula05Green] (h1) -- ++(-1.5,0) node [left] {$40$};
            \draw [latex-,line width=3pt,draw=myParula05Green] (a52) -- ++(1.5,0) node [right] {$40$};
            
        \end{scope}
    
    \end{tikzpicture}
    \caption{As a result of the split view, the vote of slot $6$ is tied---$\mathsf{Left}$ increases by roughly half of honest votes (here $40$) and $\mathsf{Right}$ increases by roughly half of honest votes (here $40$).}
    \label{fig:lmd-attack-4}
\end{figure}
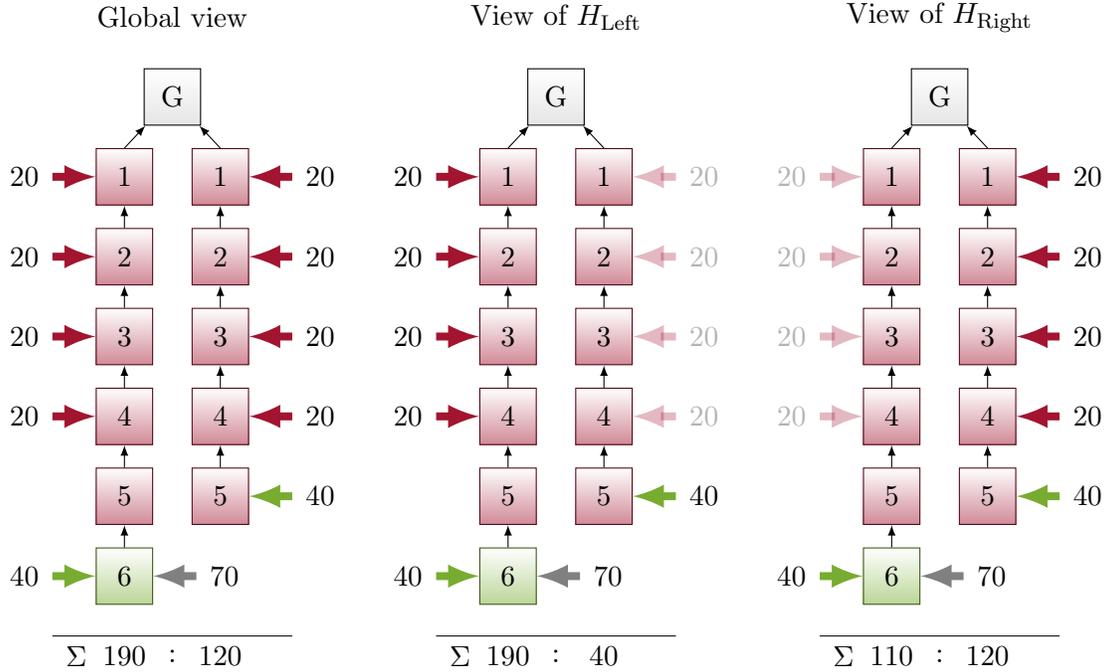

\begin{figure}
    \centering
    \begin{tikzpicture}[blockchain,x=0.75cm,scale=0.85]
    
        \begin{scope}
    
            \node at (0,1) {Global view};
            
            \node at (-1,-7) {$120$};
            \node at (1,-7) {$120$};
            \node at (-2,-7) {$\Sigma$};
            \node at (0,-7) {:};
            \draw (-2.5,-6.75) -- (2.5,-6.75);
    
            \node [block] (G) at (0,0) {G};
            
            \node [block-adv] (a11) at (-1,-1) {1};
            \node [block-adv] (a12) at (+1,-1) {1};
            \node [block-adv] (a21) at (-1,-2) {2};
            \node [block-adv] (a22) at (+1,-2) {2};
            \node [block-adv] (a31) at (-1,-3) {3};
            \node [block-adv] (a32) at (+1,-3) {3};
            \node [block-adv] (a41) at (-1,-4) {4};
            \node [block-adv] (a42) at (+1,-4) {4};
            \node [block-adv] (a51) at (-1,-5) {5};
            \node [block-adv] (a52) at (+1,-5) {5};
            
            \draw [link-adv] (a11.north) -- (G);
            \draw [link-adv] (a12.north) -- (G);
            \draw [link-adv] (a21.north) -- (a11);
            \draw [link-adv] (a22.north) -- (a12);
            \draw [link-adv] (a31.north) -- (a21);
            \draw [link-adv] (a32.north) -- (a22);
            \draw [link-adv] (a41.north) -- (a31);
            \draw [link-adv] (a42.north) -- (a32);
            \draw [link-adv] (a51.north) -- (a41);
            \draw [link-adv] (a52.north) -- (a42);
            
            \draw [latex-,line width=3pt,draw=myParula07Red] (a11) -- ++(-1.5,0) node [left] {$20$};
            \draw [latex-,line width=3pt,draw=myParula07Red] (a21) -- ++(-1.5,0) node [left] {$20$};
            \draw [latex-,line width=3pt,draw=myParula07Red] (a31) -- ++(-1.5,0) node [left] {$20$};
            \draw [latex-,line width=3pt,draw=myParula07Red] (a41) -- ++(-1.5,0) node [left] {$20$};
            \draw [latex-,line width=3pt,draw=myParula07Red] (a12) -- ++(1.5,0) node [right] {$20$};
            \draw [latex-,line width=3pt,draw=myParula07Red] (a22) -- ++(1.5,0) node [right] {$20$};
            \draw [latex-,line width=3pt,draw=myParula07Red] (a32) -- ++(1.5,0) node [right] {$20$};
            \draw [latex-,line width=3pt,draw=myParula07Red] (a42) -- ++(1.5,0) node [right] {$20$};

            \node [block-hon] (h1) at (-1,-6) {6};
            \draw [link-hon] (h1.north) -- (a51);

            \draw [latex-,line width=3pt,draw=myParula05Green] (h1) -- ++(-1.5,0) node [left] {$40$};
            \draw [latex-,line width=3pt,draw=myParula05Green] (a52) -- ++(1.5,0) node [right] {$40$};
            
        \end{scope}
    
        \begin{scope}[xshift=6cm]
    
            \node at (0,1) {View of $H_{\mathrm{Left}}$};
            
            \node at (-1,-7) {$120$};
            \node at (1,-7) {$40$};
            \node at (-2,-7) {$\Sigma$};
            \node at (0,-7) {:};
            \draw (-2.5,-6.75) -- (2.5,-6.75);
    
            \node [block] (G) at (0,0) {G};
            
            \node [block-adv] (a11) at (-1,-1) {1};
            \node [block-adv] (a12) at (+1,-1) {1};
            \node [block-adv] (a21) at (-1,-2) {2};
            \node [block-adv] (a22) at (+1,-2) {2};
            \node [block-adv] (a31) at (-1,-3) {3};
            \node [block-adv] (a32) at (+1,-3) {3};
            \node [block-adv] (a41) at (-1,-4) {4};
            \node [block-adv] (a42) at (+1,-4) {4};
            \node [block-adv] (a51) at (-1,-5) {5};
            \node [block-adv] (a52) at (+1,-5) {5};
            
            \draw [link-adv] (a11.north) -- (G);
            \draw [link-adv] (a12.north) -- (G);
            \draw [link-adv] (a21.north) -- (a11);
            \draw [link-adv] (a22.north) -- (a12);
            \draw [link-adv] (a31.north) -- (a21);
            \draw [link-adv] (a32.north) -- (a22);
            \draw [link-adv] (a41.north) -- (a31);
            \draw [link-adv] (a42.north) -- (a32);
            \draw [link-adv] (a51.north) -- (a41);
            \draw [link-adv] (a52.north) -- (a42);

            \draw [latex-,line width=3pt,draw=myParula07Red] (a11) -- ++(-1.5,0) node [left] {$20$};
            \draw [latex-,line width=3pt,draw=myParula07Red] (a21) -- ++(-1.5,0) node [left] {$20$};
            \draw [latex-,line width=3pt,draw=myParula07Red] (a31) -- ++(-1.5,0) node [left] {$20$};
            \draw [latex-,line width=3pt,draw=myParula07Red] (a41) -- ++(-1.5,0) node [left] {$20$};
            \draw [latex-,line width=3pt,draw=myParula07Red,opacity=0.3] (a12) -- ++(1.5,0) node [right] {$20$};
            \draw [latex-,line width=3pt,draw=myParula07Red,opacity=0.3] (a22) -- ++(1.5,0) node [right] {$20$};
            \draw [latex-,line width=3pt,draw=myParula07Red,opacity=0.3] (a32) -- ++(1.5,0) node [right] {$20$};
            \draw [latex-,line width=3pt,draw=myParula07Red,opacity=0.3] (a42) -- ++(1.5,0) node [right] {$20$};

            \node [block-hon] (h1) at (-1,-6) {6};
            \draw [link-hon] (h1.north) -- (a51);

            \draw [latex-,line width=3pt,draw=myParula05Green] (h1) -- ++(-1.5,0) node [left] {$40$};
            \draw [latex-,line width=3pt,draw=myParula05Green] (a52) -- ++(1.5,0) node [right] {$40$};
            
        \end{scope}
    
        \begin{scope}[xshift=12cm]
    
            \node at (0,1) {View of $H_{\mathrm{Right}}$};
            
            \node at (-1,-7) {$40$};
            \node at (1,-7) {$120$};
            \node at (-2,-7) {$\Sigma$};
            \node at (0,-7) {:};
            \draw (-2.5,-6.75) -- (2.5,-6.75);
    
            \node [block] (G) at (0,0) {G};
            
            \node [block-adv] (a11) at (-1,-1) {1};
            \node [block-adv] (a12) at (+1,-1) {1};
            \node [block-adv] (a21) at (-1,-2) {2};
            \node [block-adv] (a22) at (+1,-2) {2};
            \node [block-adv] (a31) at (-1,-3) {3};
            \node [block-adv] (a32) at (+1,-3) {3};
            \node [block-adv] (a41) at (-1,-4) {4};
            \node [block-adv] (a42) at (+1,-4) {4};
            \node [block-adv] (a51) at (-1,-5) {5};
            \node [block-adv] (a52) at (+1,-5) {5};
            
            \draw [link-adv] (a11.north) -- (G);
            \draw [link-adv] (a12.north) -- (G);
            \draw [link-adv] (a21.north) -- (a11);
            \draw [link-adv] (a22.north) -- (a12);
            \draw [link-adv] (a31.north) -- (a21);
            \draw [link-adv] (a32.north) -- (a22);
            \draw [link-adv] (a41.north) -- (a31);
            \draw [link-adv] (a42.north) -- (a32);
            \draw [link-adv] (a51.north) -- (a41);
            \draw [link-adv] (a52.north) -- (a42);

            \draw [latex-,line width=3pt,draw=myParula07Red,opacity=0.3] (a11) -- ++(-1.5,0) node [left] {$20$};
            \draw [latex-,line width=3pt,draw=myParula07Red,opacity=0.3] (a21) -- ++(-1.5,0) node [left] {$20$};
            \draw [latex-,line width=3pt,draw=myParula07Red,opacity=0.3] (a31) -- ++(-1.5,0) node [left] {$20$};
            \draw [latex-,line width=3pt,draw=myParula07Red,opacity=0.3] (a41) -- ++(-1.5,0) node [left] {$20$};
            \draw [latex-,line width=3pt,draw=myParula07Red] (a12) -- ++(1.5,0) node [right] {$20$};
            \draw [latex-,line width=3pt,draw=myParula07Red] (a22) -- ++(1.5,0) node [right] {$20$};
            \draw [latex-,line width=3pt,draw=myParula07Red] (a32) -- ++(1.5,0) node [right] {$20$};
            \draw [latex-,line width=3pt,draw=myParula07Red] (a42) -- ++(1.5,0) node [right] {$20$};

            \node [block-hon] (h1) at (-1,-6) {6};
            \draw [link-hon] (h1.north) -- (a51);

            \draw [latex-,line width=3pt,draw=myParula05Green] (h1) -- ++(-1.5,0) node [left] {$40$};
            \draw [latex-,line width=3pt,draw=myParula05Green] (a52) -- ++(1.5,0) node [right] {$40$};
            
        \end{scope}
    
    \end{tikzpicture}
    \caption{At the end of slot $6$, the proposer boost disappears. In the view of each honest validator, both chains gained roughly the same amount of votes, namely half of the honest validators' votes. Assuming a perfect split of $|H_{\mathrm{Left}}|=|H_{\mathrm{Right}}|=40$, $\mathsf{Left}$:$\mathsf{Right}$ is now $120:40$ in the view of $H_{\mathrm{Left}}$ and $40:120$ in the view of $H_{\mathrm{Right}}$ (up from $80:0$ and $0:80$, respectively). The pattern of Figures~\ref{fig:lmd-attack-1}--\ref{fig:lmd-attack-5} repeats in subsequent slots, with the honest validators in $H_{\mathrm{Left}}$ and $H_{\mathrm{Right}}$ solely voting for the chains $\mathsf{Left}$ and $\mathsf{Right}$, respectively, thus maintaining a balance of weights (in global view--in the LMD view of each validator, they keep voting for the chain they see leading, and `cannot understand' why other honest validators keep voting for the other chain) and perpetuating the adversarially induced split view.}
    \label{fig:lmd-attack-5}
\end{figure}
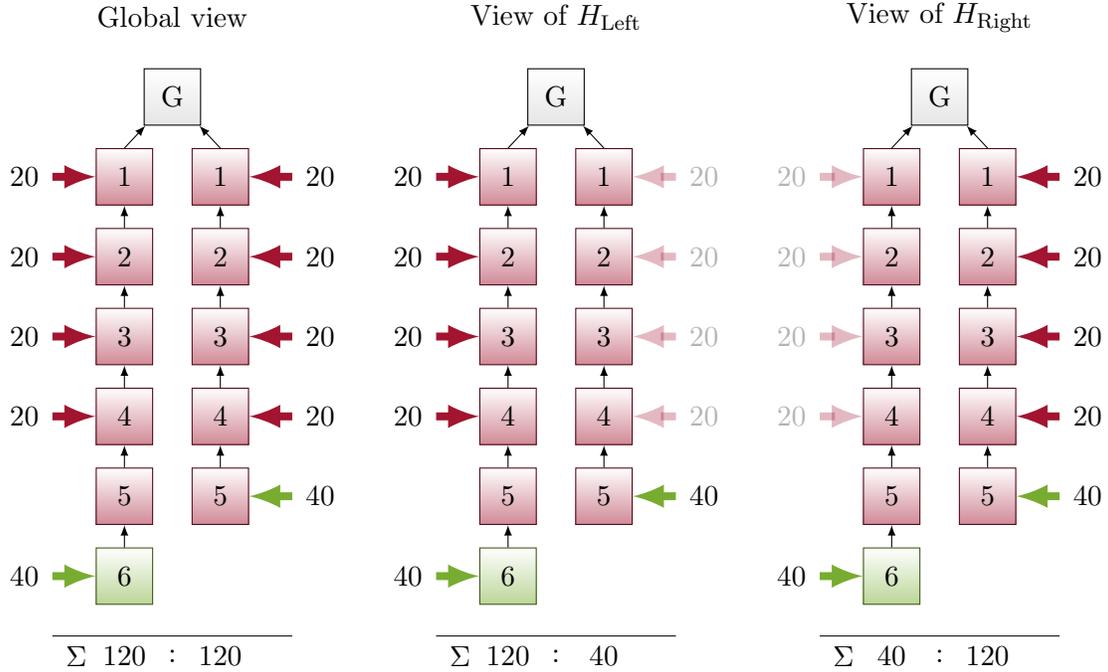

Let $W=100$ denote the number of validators per slot. 
Suppose the proposal weight is $\Wp=0.7 W = 70$, and the fraction of adversarial validators is $\beta=0.2$. 
Furthermore, for simplicity, assume that the attack starts when there are five consecutive slots with adversarial proposers.

During the first four slots, the adversary creates two parallel chains $\mathsf{Left}$ and $\mathsf{Right}$ of $4$ blocks each, which are initially kept private from the honest validators. 
Each block is voted on by the $20$ adversarial validators from its slot. 
Thus, there are equivocating votes for the conflicting blocks proposed at the same slot. 
For the fifth slot, the adversary includes all equivocating votes for the $\mathsf{Left}$ chain into a block and attaches it on the $\mathsf{Left}$ chain; and all the equivocating votes for the $\mathsf{Right}$ chain into an equivocating block and attaches it on the $\mathsf{Right}$ chain.
With this, votes are `batched' in the following sense. 
The adversary releases the two equivocating blocks from the fifth slot in such a way that roughly half of the honest validators see the $\mathsf{Left}$ block first 
(call $H_{\mathrm{Left}}$ that set of honest validators) and with it all the equivocating votes for the $\mathsf{Left}$ chain; and half of the honest validators see the $\mathsf{Right}$ block first (call $H_{\mathrm{Right}}$ that set of honest validators) and with it all the equivocating votes for the $\mathsf{Right}$ chain. 
(Note that this trick is not needed in networks with adversarial delay, where the release of equivocating votes can be targeted such that each honest validator either sees all $\mathsf{Left}$ votes first or all $\mathsf{Right}$ votes first.)
By the LMD rule, validators in $H_{\mathrm{Left}}$ and $H_{\mathrm{Right}}$ believe that $\mathsf{Left}$ and $\mathsf{Right}$ have $80$ votes, respectively. They also believe that the respective other chain has $0$ votes as the later arriving votes are not considered due to LMD (\cf Figure~\ref{fig:lmd-attack-1}).

Now suppose the validator of slot $6$ is honest and from set $H_{\mathrm{Left}}$. Then, it proposes a block extending $\mathsf{Left}$. $\mathsf{Left}$ gains a proposal boost equivalent to $70$ votes (\cf Figures~\ref{fig:lmd-attack-2} and~\ref{fig:lmd-attack-3}).
Thus, validators in $H_{\mathrm{Left}}$ see $\mathsf{Left}$ as leading with $150$ votes and vote for it. Validators in $H_{\mathrm{Right}}$ believe that $\mathsf{Left}$ has $70$ votes while $\mathsf{Right}$ has $80$ votes, so they vote for $\mathsf{Right}$. As a result, their vote is tied---$\mathsf{Left}$ increases by roughly half of honest votes and $\mathsf{Right}$ increases by roughly half of honest votes (\cf Figure~\ref{fig:lmd-attack-4}).
At the end of the slot, the proposer boost disappears. In the view of each honest validator, both chains gained roughly the same amount of votes, namely half of the honest validators' votes. Assuming a perfect split of $|H_{\mathrm{Left}}|=|H_{\mathrm{Right}}|=40$, $\mathsf{Left}$:$\mathsf{Right}$ is now 120:40 in the view of $H_{\mathrm{Left}}$ and 40:120 in the view of $H_{\mathrm{Right}}$ (up from 80:0 and 0:80, respectively).

This pattern repeats in subsequent slots, with the honest validators in $H_{\mathrm{Left}}$ and $H_{\mathrm{Right}}$ solely voting for the chains $\mathsf{Left}$ and $\mathsf{Right}$, respectively, thus maintaining a balance of weights the in global view and perpetuating the adversarially induced split view---In the LMD view of each validator, they keep voting for the chain they see leading, and `cannot understand' why other honest validators keep voting for the other chain. (\cf Figure~\ref{fig:lmd-attack-5})

\section*{Acknowledgment}
We thank
Danny Ryan, Aditya Asgaonkar, and Francesco D'Amato for feedback and fruitful discussions.
JN, ENT and DT are supported by a gift from the Ethereum Foundation.
JN is supported by the Reed-Hodgson Stanford Graduate Fellowship.
ENT is supported by the Stanford Center for Blockchain Research.

\printbibliography

\end{document}